\documentclass[fleqn,usenatbib]{mnras}

\usepackage{siunitx}

\usepackage{graphicx}	
\usepackage{amsmath}	
\usepackage{amssymb}	
\usepackage{multirow}
\usepackage{xcolor}
\usepackage{newtxtext,newtxmath}
\defcitealias{Yang2012ApJ}{Y12}
\defcitealias{OwenYang2021MNRAS}{OY22}
\defcitealias{Yang2013MNRAS}{Y13}

\title[Emission from hadronic and leptonic galaxy bubbles]{Emission from hadronic and leptonic processes in galactic {jet-driven} bubbles}

\author[Owen \& Yang]{
Ellis R. Owen$^{1,2}$\thanks{E-mail: erowen@gapp.nthu.edu.tw (ERO), hyang@phys.nthu.edu.tw (HYKY)}, 
H.-Y. Karen Yang$^{1,2,3}$
\\
$^{1}$Institute of Astronomy, National Tsing Hua University, Hsinchu, Taiwan (ROC)\\
$^{2}$Center for Informatics and Computation in Astronomy, National Tsing Hua University, Hsinchu, Taiwan (ROC)\\
$^{3}$Physics Division, National Center for Theoretical Sciences, Taipei, 106017, Taiwan (ROC)
}

\date{Accepted XXX. Received YYY; in original form ZZZ}

\pubyear{2020}

\begin{document}
\label{firstpage}
\pagerange{\pageref{firstpage}--\pageref{lastpage}}
\maketitle

\begin{abstract} 
We investigate the multiwavelength emission from hadronic and leptonic cosmic rays (CRs) in bubbles around galaxies, analogous to the \textit{Fermi} bubbles of the Milky Way. The bubbles are modeled using 3D magnetohydrodynamical (MHD) simulations, and are driven by a 0.3 Myr intense explosive outburst from the nucleus of Milky Way-like galaxies. We compute their non-thermal emission properties at different stages throughout their evolution, up to 7 Myr, by post-processing the simulations. 
We compare the spectral and spatial signatures of bubbles with hadronic, leptonic and hybrid hadro-leptonic CR compositions. These each show broadly similar emission spectra, comprised of radio synchrotron, inverse Compton and {non-thermal} bremsstrahlung components. {However, hadronic and hybrid bubbles were found to be brighter than leptonic bubbles in X-rays, and marginally less bright at radio frequencies, and in $\gamma$-rays between $\sim$ 0.1 and a few 10s of GeV,} with a large part of their emission being driven by secondary electrons formed in hadronic interactions. {Hadronic systems were also found to be slightly brighter in high-energy $\gamma$-rays than their leptonic counterparts, owing to the $\pi^0$ decay emission that} {dominates their emission between energies of 100s of GeV and a few TeV}. 
\end{abstract}
\begin{keywords}
Cosmic rays -- galaxies: nuclei -- MHD -- gamma-ray: galaxies -- radio continuum: galaxies -- X-rays: galaxies
\end{keywords}


%

\section{Introduction} 
\label{sec:intro} 

The \textit{Fermi} bubbles are giant diffuse $\gamma$-ray structures extending roughly symmetrically above and below the Galactic Center (GC) of our Galaxy, reaching heights of around \ang{50} from the Galactic plane. Their total $\gamma$-ray power is estimated to be $>10^{37}{\rm erg}~{\rm s}^{-1}$ 
  between $(1 -  100)~{\rm GeV}$, 
  with an apparent spectral cut-off at $\sim 110$~GeV~\citep{Ackermann2014ApJ}.   
The discovery and detection of these structures more than a decade ago~\citep{Su2010ApJ, Dobler2010ApJ, Ackermann2014ApJ} attracted substantial attention, opening up many new questions about their physical nature and origin.
The subsequent detection of galactic-scale emission structures at energies above X-rays in a number of external galaxies, 
  e.g. M31 \cite[see][]{Pshirkov2016MNRAS}, 
  the Circinus galaxy~\citep{Hayashida2013ApJ} 
  and the starburst galaxies 
  NGC~3097 \citep{Li2019ApJ}, 
  NGC~253 \citep{Acero2009Sci,Abramowski2012ApJ} and M82 \citep{VERITAS2009Nat} 
  spurred the consideration that the \textit{Fermi} bubbles may not
  be particularly unusual or unique, and that 
  these phenomena could be ubiquitous throughout the galaxies of the Universe,
  perhaps even constituting a subset of 
  structures we are more familiar with in other wavebands or at different stages of their development (see also~\citealt{OwenYang2021MNRAS}, hereafter~\citetalias{OwenYang2021MNRAS}), or resulting from different ambient conditions~\citep{Zhang2021ApJ}.

Several models have been introduced to address the origin, composition and emission properties of the \textit{Fermi} bubbles \citep[see a review by][]{Yang2018Galax}. \citealt{Yang2012ApJ} (hereafter \citetalias{Yang2012ApJ}) hypothesized an origin in an intense explosive outburst of the central black hole at the GC of the Milky Way, {Sgr A*} around 1.2 Myr ago (with the \textit{Fermi} bubbles then emerging as relic cocoons of previous active galactic nucleus, AGN, activity).\footnote{Such an event would not be unprecedented, and may share certain similarities with more recent less-energetic outbursts associated with {Sgr A*}, as is widely interpreted as the explanation for time-variable Fe K$\alpha$ emission in Sgr B2~\citep[e.g.][]{Rogers2021ApJ}.} 

In this scenario, the present $\gamma$-ray emission {from the \textit{Fermi} bubbles} arises predominantly through inverse Compton scattering of an energetic non-thermal cosmic ray (CR) electron population in the remnant structures with ambient radiation supplied by the interstellar radiation field (ISRF) and the cosmological microwave background (CMB). A sub-dominant component due to non-thermal bremsstrahlung may also be present, emitted primarily from regions of high gas and non-thermal electron density within the bubble (\citetalias{OwenYang2021MNRAS}). Models of this nature are broadly referred to as `leptonic' models (see also~\citealt{Su2010ApJ, Zubovas2011MNRAS, Su2012ApJ, Fujita2013ApJ} for similar configurations). Other similar approaches invoking Sgr A* activity, but where the CR composition is not specifically required to be leptonic, invoke a pair of jet-driven outflowing bubbles assuming constant AGN activity or continuous energy injection over Myr timescales~\citep[e.g.][]{Zhang2020ApJ}. Notably, these models have been able to account for the bi-conical X-ray structures observed near the GC as part of the same phenomenon as the \textit{Fermi} bubbles. 
Alternative proposals have also been discussed, where the bubbles arise from the confluence of a number of processes operating more gradually within the inner part of the Milky Way~\citep{Thoudam2013ApJ}. These could include tidal disruption events (TDEs) occurring at regular intervals of 10s to 100s kyr~\citep{Cheng2011ApJ, Ko2020ApJ}, or the action of a bi-polar galactic outflow driven by the ongoing intense GC star-formation activity and/or the processes associated with Sgr A*~\citep{Lacki2014MNRAS}, with the resulting $\gamma$-ray glow instead arising from a hadronic CR population interacting with an advected supply of entrained gas in the wind (the `hadronic' models -- see ~\citealt{Crocker2011PhRvL, Crocker2014ApJ, Crocker2015ApJ, Mou2014ApJ, Mou2015ApJ, Cheng2014ApJ, Cheng2015ApJ, Razzaque2018Galax}). 

Both leptonic and hadronic scenarios have drawn on certain support from observations. 
For example, before the \textit{Fermi} bubbles had even been detected in $\gamma$-rays, their location was known to be X-ray dim. 
This was thought to be due to an under-dense medium, perhaps indicative of a wind~\citep{BlandHawthorn2003ApJ}. This interpretation would lean towards a hadronic composition due to the slow cooling timescales of protons compared to electrons. 
Conversely, a microwave `haze' of hard spectrum above and below the GC, would lend support to a leptonic composition~\citep{Ackermann2014ApJ}. This had been attributed to synchrotron emission from a hard spectrum of non-thermal electrons -- again, long before the $\gamma$-ray detection of the \textit{Fermi} bubbles~\citep{Finkbeiner2004ApJ}, but with later confirmation by \textit{Planck} (\citealt{Planck2013AA}; see also~\citealt{Dobler2008ApJ, Dobler2012ApJ}). Polarized radio observations at 2.3 GHz in the same regions were initially accounted for using a hadronic bubble composition~\citep{Carretti2013Natur}, however it was later shown by~\citealt{Yang2013MNRAS} (hereafter~\citetalias{Yang2013MNRAS}) that a leptonic model could be consistent with the polarization signatures too.

In recent years, a need for a clear discriminating signature between a leptonic and hadronic bubble compositions has thus emerged. This has motivated the search for TeV $\gamma$-rays and neutrinos associated with the \textit{Fermi} bubbles, which could be used as a \textit{smoking gun} for hadronic processes.\footnote{These would result from the decay of charged and neutral pions formed in hadronic interactions~\citep[e.g.][]{Dermer2009_book, Owen2021MNRAS}.} Earlier works discussed a possible connection between the $\gamma$-ray emission and neutrino detections from the direction of the bubbles~\citep{Aartsen2014PhRvL}, and it was shown by~\citet{Lunardini2013PhRvD} that, for a primary CR proton cut-off energy at or above 10 PeV, the \textit{Fermi} bubbles could account for up to $\sim$4-5 of the 28 neutrino events that had been detected above 30 TeV by IceCube by that time.\footnote{ \citet{Lunardini2014arXiv} later revised this estimate to $\sim$6-7 of 37 neutrino events in a subsequent analysis.} \cite{AdrianMartinez2014EPJC} further found 16 neutrino events coming from the direction of the \textit{Fermi} bubbles compared to 33 events in off-bubble zones. However, this would only correspond to a non-significant 1.2$\sigma$ excess from the bubbles\footnote{This is comparable to a 1.5$\sigma$ excess reported by the ANTARES neutrino telescope using track-like event signatures -- see~\citealt{Hallmann2017JPhCS}.} 
and, more recently, \cite{Fang2017PhRvD} (see also~\citealt{Sherf2017ApJ}) reported no evidence for neutrino emission originating from the \textit{Fermi} bubbles, with previously suggested IceCube excesses being consistent with the diffuse astrophysical background. Further to this, the non-detection of the \textit{Fermi} bubbles in $\gamma$-rays between 1-100 TeV with HAWC~\citep{Abeysekara2017ApJ, Fang2017PhRvD, Sherf2017ApJ} constrains the possible hadronic activity in the bubble,\footnote{Such constraints have been further strengthened by non-detections above 1 TeV with H.E.S.S.~\citep{Moulin2021arXiv} at the base of the \textit{Fermi} bubbles, and spectral a cut-off of around 500 GeV~\citep{Herold2019AA}.} although does not rule-out hybrid lepto-hadronic models~\citep[e.g. as that in][]{AlvarezHurtado2019ICRC}. 

While these observations can impart certain constraints on the composition of our own Galaxy's \textit{Fermi} bubbles, 
their relevance to galactic bubbles and/or AGN jets in other galaxies (where conditions may be very different) is unclear, and it is important to understand the observational signatures that would be associated with different bubble compositions throughout their evolution{, and the degree to which these can be used to diagnose different bubble compositions}.
In this paper, we consider the broadband multiwavelength {non-thermal} emission signatures for galaxy-scale bubbles with leptonic, hadronic and hybrid lepto-hadronic CR compositions, 
and investigate how these signatures would evolve as the system ages. This is an extension of our previous work  (\citetalias{OwenYang2021MNRAS}), where the emission signatures from ageing leptonic bubbles were {investigated, and is the first study to explore the multi-wavelength non-thermal emission signatures of hadronic and leptonic CRs in galaxy bubbles through their evolutionary progression. In this paper, we put focus on 
hadronic and leptonic emission signatures from galaxy bubbles. To do this, we invoke a baseline model motivated by the Galactic \textit{Fermi} bubbles as a stage to model the bubble emission spectra. However, we leave detailed study of dependencies on physical bubble parameters (e.g. AGN activity and variations in total energetics, host galaxy properties and the bubble environment) to future work.\footnote{{Different model parameter choices could yield galaxy bubbles with substantially modified characteristics. For example,~\citet{Yang2022NatAs} used broadly the same model configuration as~\cite{Yang2012ApJ}, but with a reduced jet duration and CR energy density. This led to a more slowly-expanding bubble which, when considered in the context of the Galactic bubbles, suggested a current age of more than twice that previously estimated by~\citealt{Yang2012ApJ} (based on the spatial extent and spectrum observed for the \textit{Fermi} bubbles, \textit{eROSITA} bubbles and Galactic microwave haze, and considering these as part of the same scenario). Here, to maintain consistency and comparability with earlier works, we retain the earlier model parameter choices of~\citealt{Yang2012ApJ} (also used in~\citetalias{OwenYang2021MNRAS}).}} We consider our approach sufficient to illustrate the expected key spectral signatures for bubbles at different evolutionary stages and for different CR compositions.}

We arrange this paper as follows: In section~\ref{sec:section2} we summarize our numerical model, initial conditions, relevant particle interactions and multiwavelength emission mechanisms. In section~\ref{sec:section3}, we present our results, including the spatial and spectral emission properties of evolving bubbles under hadronic, leptonic and hybrid composition scenarios. We discuss our findings in section~\ref{sec:section4}. Our conclusions are given in section~\ref{sec:section5}.
 
\section{Methodology}
\label{sec:section2}

\subsection{Numerical approach}
\label{sec:numerical_method}

To model bubbles emerging from galaxies, 
we perform 3D magneto-hydrodynamical (MHD) simulations, where the simulation set-up is identical to that adopted in~\citetalias{OwenYang2021MNRAS}. We refer the reader to this earlier work for a detailed description of our code, method, initial conditions and numerical techniques, and provide a brief summary of the setup and parameters here. In our simulations, the bubble inflation is driven by an initial 0.3 Myr injection {of CR energy, magnetic field energy and thermal gas (supplying kinetic and thermal energy)} from the center of the model galaxy, corresponding to an intensive explosive outburst from a period of AGN activity, with total injected energy $E_{\rm j} = 3.13\times 10^{57} \;\! {\rm erg}$ per jet.\footnote{{Note that longer outburst periods could be adopted to model systems with ongoing AGN activity~\citep[see][]{Yang2019ApJ}. In our set-up this short 0.3 Myr choice leads to the formation of remnant bubbles.}} The subsequent development of the bubble is simulated using FLASH4~\citep{Fryxell2000ApJS, Dubey2008PhST}, an adaptive mesh refinement code, in a simulation box of 75 kpc on each side.
We use a progressive grid refinement process, undertaken in parts of the simulation grid where the CR energy density exceeds $10^{-10}~{\rm erg}\;\!{\rm cm}^{-3}$ in each simulation step, with a resulting resolution ranging from 4.7 kpc (coarsest) to 0.6 kpc (finest).  We set diode boundary conditions which allow for the outflow of gas and CRs, but prevent inflows into the simulation domain. We solve the MHD equations using the directionally unsplit staggered mesh solver~\citep{Lee2009JCoPh, Lee2013JCoPh} which ensures divergence-free magnetic fields. 
In our simulations, we model the CRs as a fluid (alongside the thermal gas), for which the CR pressure evolution is solved directly. We self-consistently model dynamical coupling between the CRs and the thermal gas, where the CRs are advected with the gas. Anisotropic CR diffusion along magnetic field lines is also included, where we assume that CR diffusion perpendicular to the local magnetic field vector is negligible (as, typically, $\kappa_{\perp} \ll \kappa_{||}$ -- see, e.g.~\citealt[][]{Ensslin2003AA}), and where we adopt a value of $4.0\times 10^{28}\;\!{\rm cm}^2\;\!{\rm s}^{-1}$ for the parallel diffusion coefficient, $\kappa_{||}$. We initialize a halo magnetic field 
with a correlation length of $\ell_c = 1$ kpc
and an average strength of 1 $\mu$G~\citep{Brown2010ASPC} 
throughout the simulation domain, such that the results presented in this work correspond to `Run A' of~\citetalias{OwenYang2021MNRAS}.\footnote{\citetalias{OwenYang2021MNRAS} showed that changing the correlation length of the initial halo magnetic field only influenced the substructure of the bubble magnetic field. It did not bear any significant impact on the overall amount of emission from the resulting bubble, or its spectrum.} 
{For completeness, we report the 
evolution of the total energy and its constituent components in our simulations in Appendix~\ref{sec:appendixb}. Moreover, plots and discussion of the evolution of simulated MHD quantities in various parts of the bubbles are provided in~\citetalias{OwenYang2021MNRAS} (in particular, their Figure 1). The same simulation results (namely, their `Run A') were used as a basis for this work, and the reader is thus referred to~\citetalias{OwenYang2021MNRAS} for more detailed discussion.}   
Distinction between leptonic and hadronic CRs is not made within the simulation, and related CR cooling/heating processes, interactions and energy spectra, are neglected. These are instead considered in post-processing computations, 
which we perform on our simulation outputs. 

\subsection{Post-processing simulations}
\label{sec:post_proc}

We post-process our simulation outputs to compute emission spectra and spatial emission maps over broad energy bands 
for varying CR proton/electron compositions. {The non-thermal emission spectrum is modeled at each point of the output simulation grid (this is similar to the approach adopted in, e.g.~\citealt{Yang2013MNRAS}). 
The CR energy spectrum driving this emission is initially assumed to be a power-law in energy, with its normalization being set by local values of $e_{\rm CR}$. 
Our simulations that have been performed in the limit where CRs are not important to the overall dynamics of the bubble evolution (instead, the bubbles are dominated by kinetic and thermal energy; see Figure~\ref{fig:total_energy_simulation}) and, as such, there
is some degeneracy between the non-thermal luminosity of the bubbles and the absolute energy content in CRs.\footnote{{If the CRs were dynamically important, they would have a strong bearing on the evolution of the bubble and such a degeneracy would not arise.}} Presently, the only available constraint for the 
non-thermal luminosity of any known Galaxy bubbles is that provided by the \textit{Fermi} bubbles of the Milky Way. We thus introduce a tuning parameter, $f_{\rm emit}$, to explicitly set the non-thermal luminosity of our simulated bubbles to match this constraint, which practically acts to scale the CR energy density down. 
In line with earlier work that matched a similarly-configured numerical simulation to the non-thermal spectrum of the \textit{Fermi} bubbles, we adopt a value of $f_{\rm emit}=0.003$~\citep{Yang2017ApJ}. {In our treatment, alternative choices of this value would act to linearly re-scale the non-thermal emission from the bubbles.}}

While this post-processing approach cannot strictly capture the true spectral evolution of the CRs\footnote{In particular the CR electrons, which have a much shorter cooling timescale than their counterpart protons at similar energies.} throughout the simulation domain (for which more sophisticated spectral tracking methods are required, as that invoked in  e.g.~\citealt{Yang2017ApJ}),{ it provides a good approximation with substantially reduced computational requirements. This reduced complexity brings bubble simulations covering large spatial volumes over long evolutionary timescales into computational reach. We consider our approach to be a good first-order approximation} because the dynamical time of expansion of the bubble is always shorter than the radiative cooling time estimated from our simulations (except for the very early stage of the bubble evolution in the case of CR electrons, which is accounted for in our treatment of spectral ageing -- cf. section~\ref{sec:leptonic_bubbles}). 

\subsection{CR interactions and emitted radiation}
\label{sec:cr_spec_int}

\subsubsection{Leptonic model}
\label{sec:leptonic_bubbles}

In the leptonic scenario, we consider a purely electronic CR composition.\footnote{We refer to electrons and positrons together as electrons, as the distinction is inconsequential for this work. The short survival times of other heavier leptonic species (e.g. muons, with a lifetime of $\tau_{\rm \mu^{\pm}} = 2.2 \mu\text{s}$ -- see~\citealt{Tanabashi2018PRD}) makes them implausible as a dominant CR component or important in terms of radiated emission from within a galaxy bubble, so these are not included in our model.} 
While we do not explicitly track the evolution of the electron spectrum due to electron cooling during the simulation, we account for this effect by emulating the evolution found by~\citealt{Yang2017ApJ}. 
We adopt a simple power-law for the \textit{initial}, injected CR electron differential energy spectrum, given by:
\begin{equation}
\frac{{\rm d}n_{\rm e}}{{\rm d}\gamma_{\rm e}} = \mathcal{N}_e^{\rm lep} \left(\frac{\gamma_{\rm e}}{\gamma_{\rm 0,e}}\right)^{-s_{\rm e}} \ ,
\label{eq:electron_spectrum}
\end{equation}
where 
$n_{\rm e}$ is the CR electron number density, 
$\gamma_{\rm e}$ is the electron Lorentz factor (which is related to the electron energy by $E_{\rm e} = \gamma_{\rm e} \;\! m_{\rm e} c^2$), 
$\gamma_{\rm 0,e}$ specifies the lower energy cut-off of the CR electron spectrum, 
$s_{\rm e}$ is the power-law spectral index, and 
the normalization $\mathcal{N}_e^{\rm lep}$ is set by the CR energy density $e_{\rm CR}$: 
\begin{equation}
\mathcal{N}_e^{\rm lep} = f_{\rm emit}\;\!e_{\rm CR}\;\!\begin{cases}
\frac{\gamma_{\rm 0,e}^{-s_{\rm e}} }{\ln (\gamma_{\rm 1,e}/\gamma_{\rm 0,e})} \hspace{2.3cm} \text{if}\;\!s_{\rm e} = 2 \ , \\
\frac{(2-s_{\rm e}) \gamma_{\rm 0,e}^{-s_{\rm e}}}{\gamma_{\rm 1,e}^{2-s_{\rm e}} - \gamma_{\rm 0,e}^{2-s_{\rm e}}} \hspace{2.3cm} \text{if}\;\!s_{\rm e} \neq 2 \ .
\end{cases}
\label{eq:electron_norm}
\end{equation}
Here, $\gamma_{\rm 1, e}$ is introduced as the upper spectral cut-off. We adopt a fiducial value for the spectral index of $s_{\rm e} = 2$, but consider that other similar, reasonable choices would be no less physical and note that the exact choice (if reasonable) does not strongly affect the results of this work. For the lower spectral limit, we adopt a value of $E_{\rm min} = \gamma_{\rm 0,e} m_{\rm c} c^2 = 1$ GeV, as we consider electrons at energies below this would undergo rapid loses in the early stages of the bubble evolution, during the first 0.1 Myr, due to Coulomb collisions in dense regions near the center of the host galaxy.\footnote{This choice is appropriate for a galactic center gas density of around 500~${\rm cm}^{-3}$ initially encountered by the CRs, which is broadly consistent with estimated mean gas densities for the Milky Way and other galaxies~\citep[e.g.][]{Oka2005ApJ, Heiner2013MNRAS, Mills2018ApJ}.} For $E_{\rm max} = \gamma_{\rm 1,e} m_{\rm c} c^2$, we adopt a value of $10$ TeV (following, e.g.~\citealt{Yang2017ApJ}). 

Our initial spectrum, given by equation~\ref{eq:electron_spectrum} is evolved to an appropriate time $t$ following the method in~\citetalias{OwenYang2021MNRAS}, which solves the electron kinetic equation (e.g. ~\citealt{Kardashev1962SvA}) with cooling due to radiative losses (synchrotron cooling in ambient magnetic fields, and inverse Compton cooling in the CMB, and the ISRF near the host galaxy) and adiabatic losses. {To approximate these radiative and adiabatic losses in a way that reasonably captures the conditions experienced by a population of CR electrons propagating through the diverse conditions of a galaxy bubble, we 
use cooling timescales determined for tracer particles in the simulations of \citet{Yang2017ApJ} (see their Figure 2), where a very similar simulation configuration to the present work was adopted (but where the CR electron spectral evolution had additionally been self-consistently modeled). Cooling rates averaged over an ensemble of tracers (to reflect a range of different particle tracks through the developing bubble) were scaled to an appropriate energy from the averaged energy of the tracer particles, and interpolated/extrapolated to the required simulation time.} The resulting spectrum deviates from a power-law only above a few {hundred} GeV, and leads to the emergence of a natural spectral cut-off of around 1 TeV after $\sim 0.2$ Myr. While the spectral evolution is very severe at high energies due to rapid cooling during the initial stages of the bubble expansion, its evolution is inconsequential after $\sim 0.5$ Myr. {We refer the reader to Figure 5 of \citetalias{OwenYang2021MNRAS}, which shows the {(normalized)} spectral evolution of the CRs in our leptonic model.} {We note that the spectral evolution is performed on equation~\ref{eq:electron_spectrum} after the spectral normalization has already been computed. This ensures that cooling is more self-consistently modeled as an evolution of an initial spectrum, and avoids excessive energy being fed into the lower energy component of the CR electron spectrum (as would result if the spectral normalization were performed after accounting for its ageing/cooling).}

The CR electrons emit radiation as they cool. Their spectral emissivity may be calculated as the sum of 
the emission contributions from all relevant processes:
\begin{align}
j(\epsilon) & = \frac{{\rm d}\epsilon}{{\rm d}t\;\!{\rm d}\nu\;\!{\rm d}V}  \nonumber \\
&= j_{\rm sy}(\epsilon) + j_{\rm ic}(\epsilon) +j_{\rm ff}^{\rm nt}(\epsilon) \ ,
\label{eq:spec_emissivity_lep}
\end{align}
 i.e. synchrotron, inverse Compton {and non-thermal bremsstrahlung} (for details, we refer the reader to~\citetalias{OwenYang2021MNRAS}\footnote{{\citetalias{OwenYang2021MNRAS} included an additional emission component from thermal bremsstrahlung. As we focus on the non-thermal emission properties in the current work, this component is not included. 
 We find that the thermal bremsstrahlung emission obtained by our post-processing approach is dependent on the simulation resolution and the age/size of the bubble. 
 To properly resolve the thermal emission from galaxy bubbles, future work would require substantially higher resolution simulations than used here.
 }}). Here we introduce $\epsilon = h\nu/m_{\rm e}c^2$ as the dimensionless photon energy (normalized to the rest mass of an electron), $h$ as the Planck constant and $\nu$ as the photon frequency. In the case of inverse Compton emission, we adopt a target radiation field comprised of CMB photons, and an ISRF associated with the host galaxy, based on a simple 3-component model, following that suggested for `normal' galaxies by~\citet{Chakraborty2013ApJ} (see also the optical and infrared ISRF contributions in \citealt{Cirelli2009NuPhB}). This is specified in~\citetalias{OwenYang2021MNRAS}.  In our post-processing method, we use equation~\ref{eq:spec_emissivity_lep} at each point throughout our simulation grid, and compute the total emitted bubble spectrum by summing this over the required emission volume.

\subsubsection{Hadronic model}
\label{sec:hadronic_bubbles}

In the hadronic scenario, we consider a CR energy density comprised entirely of protons, with a spectrum following the same parametrised form as was adopted for electrons in the leptonic scenario, i.e.
\begin{equation}
\frac{{\rm d}n_{\rm p}}{{\rm d}\gamma_{\rm p}} = \mathcal{N}_p^{\rm had} \left(\frac{\gamma}{\gamma_{\rm 0,p}}\right)^{-s_{\rm p}}  \ .
\label{eq:proton_spectrum}
\end{equation}
Here, $n_{\rm p}$ is the CR proton number density, $\gamma_{\rm p}$ is the proton Lorentz factor (related to proton energy by $E_{\rm p} = \gamma_{\rm p} m_{\rm p} c^2$), and $s_{\rm p}$ is the power-law spectral index, which we set to be equivalent to that for electrons, i.e. $s_{\rm p} = s_{\rm e} = 2$, as we would expect the mechanism accelerating the charged particles to relativistic energies to be the same. $\gamma_{\rm 0,p} = E_{\rm min}/m_{\rm p} c^2$ and $\gamma_{\rm 1,p}= E_{\rm max}/m_{\rm p} c^2$ set the lower and upper spectral limits, where $E_{\rm min}$ and $E_{\rm max}$ retain the same definitions and values as adopted in the leptonic model. The normalization term $\mathcal{N}_p^{\rm had}$ also follows the same form as that initially used for the electrons:
\begin{equation}
\mathcal{N}_p^{\rm had} = f_{\rm emit}\;\!e_{\rm CR}\;\! \begin{cases}
\frac{\gamma_{\rm 0, p}^{-s_{\rm p}} }{\ln (\gamma_{\rm 1, p}/\gamma_{\rm 0, p})} \hspace{2.3cm} \text{if}\;\!s_{\rm p} = 2 \ , \\
\frac{(2-s_{\rm p}) \gamma_{\rm 0, p}^{-s_{\rm p}}}{\gamma_{\rm 1, p}^{2-s_{\rm p}} - \gamma_{\rm 0, p}^{2-s_{\rm p}}} \hspace{2.3cm} \text{if}\;\!s_{\rm p} \neq 2 \ .
\end{cases}
\label{eq:proton_norm}
\end{equation}
Unlike the leptonic model, the cooling time for protons is substantially longer than that for electrons (under comparable conditions). As such, spectral ageing is unlikely to be important for the CR protons, and we adopt the spectral form given by equation~\ref{eq:proton_spectrum} at all times, with no spectral evolution.

For the hadronic model, emission from galaxy bubbles is {only partly} associated with the CR protons {directly (mainly through the decay of neutral pions).} {Much of it is instead} mediated by the production of secondary electrons (via the decay of charged pions) through hadronic interactions.  
For the conditions expected in a galaxy bubble, the most important hadronic interaction is the proton-proton (pp) process.\footnote{While p$\gamma$ interactions may also arise between CR protons and the photons of ambient radiation fields, we estimate that the pp interaction would dominate in galaxy bubble conditions by a few orders of magnitude.} This proceeds when a high-energy proton interacts with a low energy proton or nucleus in the thermal plasma inside a galaxy bubble. It can occur above a threshold proton kinetic energy of $T_{\rm p}^{\rm th}/c^2 = 2m_{\pi^0} + m_{\pi^0}^2/2m_{\rm p} = 0.28~\text{GeV}/c^2$ 
at a rate given by: 
    \begin{equation}
        \label{eq:rate_pp}
        \dot{n}_{\rm p\pi}(\gamma_{\rm p}) \;\! {\rm d}\gamma_{\rm p}= n_{\rm H} \;\!{c}\;\!\sigma_{\rm p\pi}(\gamma_{\rm p})  \;\! n_{\rm p}(\gamma_{\rm p}) \;\! {\rm d}\gamma_{\rm p} \ , 
    \end{equation}
    \citep[e.g.][]{Owen2018MNRAS}, 
where $n_{\rm p}$ is the CR proton density, 
and ${\sigma}_{\rm p\pi}$ is the total inelastic pp interaction cross section, which may be parameterised as
\begin{equation}%
\label{eq:pp_cs}%
   \sigma_{\rm p\pi} = \left( 30.7 - 0.96\ln(\chi) + 0.18(\ln\chi)^{2} \right)\left( 1 - \chi^{-1.9}   \right)^{3}~\rm{mb} 
\end{equation}%
   \citep{Kafexhiu2014}. Here, $\chi = T_{\rm p}/T_{\rm p}^{\rm{th}}$, for $T_{\rm p}^{\rm th} = (\gamma_{\rm p}^{\rm{th}}-1)\;\!m_{\rm p}c^2$ is introduced as the threshold proton kinetic energy, below which the interaction does not arise. 
   The energy of the interaction is completely dominated by the energy of the interacting CR proton, and it proceeds as a two-step process: first, resonance baryons are formed via $\rm{p}\rm{p} \rightarrow \rm{p}  \Delta^{+~\;}$ or $\rm{p}\rm{p} \rightarrow \rm{n} \Delta^{++}$~\citep{Almeida1968PR, Skorodko2008EPJA}, which then decay rapidly (on timescales of $5.63\times 10^{-24}$ s -- see~\citealt{Patrignani2016ChPh}) to yield charged and neutral pions, with energy-dependent multiplicities of the three pion species \citep[see, e.g.][]{Jain1993physrep}.
  The energy-dependence of the pion multiplicities varies most strongly between the threshold energy and $T_{\rm p} \sim 50$~GeV
  \citep{Almeida1968PR, Blattnig2000, Skorodko2008EPJA},
changing from a production ratio of $\{\pi^+, \pi^-, \pi^0\} = \{ 0.6, 0.1, 0.3\}$ 
  at 1~GeV, and stabilizing at around 
  $\{ 0.3, 0.4, 0.3\}$ by 50~GeV~\citep{Jacobsen2015}. The production of secondary electrons is regulated by the decay of the charged pions, while their spectrum is determined by both the energetics of the pp interaction and the local cooling rate experienced by the electrons. The 
  multiple electron secondaries produced in a given interaction would  
have a roughly equal share of energy.\footnote{This follows from the presence of a strong peak in 
the differential electron production cross section in the pp interaction~\citep{Murphy1987ApJS, Berrington2003ApJ}, which indicates that many of the electrons would be produced at similar energies.} When accounting for losses to neutrinos, the secondary electrons characteristically inherit a few percent of the energy of the primary proton~\citep[e.g.][]{Owen2018MNRAS}. 

{The cooling timescales of the secondary electrons within the bubbles could be substantial -- particularly during the later evolutionary stages of the system, when they can even exceed a bubble's age at low energies. Strictly, therefore, a treatment should be adopted which self-consistently models the secondary electron spectra at different stages of a bubble's evolution. Following~\cite{Kardashev1962SvA}, and considering the limit where adiabatic losses are inconsequential (which was shown in~\citealt{Yang2017ApJ} to be reasonable for much of the bubble's lifetime), the electron spectrum would exhibit two regimes, separated by a spectral cooling break, i.e:}
\begin{equation}
    {n_{\rm e}(\gamma_{\rm e}, t) \approx \left\{\begin{array}{lr}
        \Gamma_{\rm e}(\gamma_{\rm e})\;\!t, & \text{for } t\ll t_{\rm cool}(\gamma_{\rm e})\\
        \Gamma_{\rm e}(\gamma_{\rm e})\;\!t_{\rm cool}(\gamma_{\rm e}), & \text{for } t\gg t_{\rm cool} (\gamma_{\rm e})
        \end{array} \right.\; .}
        \label{eq:kard_sol}
\end{equation}
{Here, $t$ is the age of the bubble,  $t_{\rm cool}$ is the energy-dependent electron cooling timescale and $\Gamma_{\rm e}$ is the volumetric injection rate of secondary CR electron particles. Thus, for electrons of energies such that $t_{\rm cool} (\gamma_{\rm e}) \ll t$, a steady-state spectrum would be attained, whereas electrons of lower energies accumulate over time.}

{In this work, we are primarily concerned with the emission signatures from CR electrons and proton populations in galaxy bubbles. These are dominated by non-thermal radiative processes, in particular synchrotron and inverse Compton emission. Only electrons of relatively high energies are strongly involved with these emission processes, which would practically be dominated by CR electrons of energies above} 
\begin{equation}
    {E_{\rm e} \approx 10.0 \;\! \left( \frac{\nu}{100~{\rm GHz}} \right)^{1/2} \;\! \left( \frac{B}{100~{\rm \mu G}} \right)^{-1/2} \;\! {\rm GeV} }\ ,
\end{equation}
{for synchrotron emission of frequency $\nu$ in a magnetic field of strength $B$, or }
\begin{equation}
    {E_{\rm e} \approx 300.0 \;\! \left( \frac{E_{\gamma}}{0.1\;\!{\rm GeV}} \right)^{1/2} \;\! \left( \frac{T_{\rm rad}}{1\;\!{\rm K}} \right)^{-1/2} \;\! {\rm GeV}} \ ,
\end{equation}
{for inverse Compton emission, with scattered photons of energy $E_{\gamma}$, and a target thermal radiation field of temperature $T_{\rm rad}$~\citep[e.g.][]{Crocker2007ApJ}.
Over the range of conditions in our simulations, particularly in regions of higher CR energy density which would dominate the emission volume, we estimate the non-thermal emission in the hadronic model is dominated by secondary electrons of energies above 
a few 10s of GeV.}

{Radiative cooling timescales for CR electrons of these energies can be estimated from the results of \cite{Yang2017ApJ}, which used tracer particles to track the position and energies of 
CRs through their simulations. As their simulation set-up was very similar to that used here, we consider the cooling timescales estimated for tracer particles 
located towards the edge of the bubble (and thus falling within the main non-thermal emission volume for much of the bubble's lifetime) would be representative for our purposes.\footnote{{\cite{Yang2017ApJ} computed cooling timescales up to 1.2 Myr.  \citetalias{OwenYang2021MNRAS} found bubble conditions do not vary rapidly after the first 1 Myr, so these cooling timescales would not increase substantially at later times.}} By scaling their results, electron cooling times would be $t_{\rm cool}\sim 0.6\;\!{\rm Myr}$ at 250 GeV, or $t_{\rm cool}\sim 3\;\!{\rm Myr}$ at 50 GeV.
Thus, we may consider that the non-thermal emission from the hadronic bubble is dominated by secondary electrons with a timescale shorter than (or, at most, comparable to) the lifetime of the bubble over most of its evolution. Thus, in equation~\ref{eq:kard_sol}, they would typically fall within the steady-state regime.}

{A further assumption that is made in equation~\ref{eq:kard_sol} is that the volumetric injection term $\Gamma_{\rm e}(\gamma_{\rm e})$ of secondary CR electrons is constant over the electron cooling timescale (if taking $t_{\rm cool}<t$). Indeed, in our post-processing approach, the time-evolution of this quantity is not even accessible. As the thermal kinetic properties and hydrodynamical structure of the galaxy bubbles can evolve over timescales shorter than this, it cannot be strictly guaranteed that the volumetric secondary CR injection rate would not vary. As such, our approximation of a constant injection term over an electron cooling timescale must be assessed. 
The volumetric interaction rate is practically set by the product of the local gas density and CR energy density. As these quantities are co-evolved in our simulations (with the exception of minimal redistribution of CR energy density by diffusion)},\footnote{{Taking $\kappa_{||} = 4.0\times 10^{28}\;\!{\rm cm}^2\;\!{\rm s}^{-1}$, diffusion would typically redistribute CRs over length-scales of $\sim 2$ kpc over a lifetime of $\sim 10$ Myr, compared to a bubble size of 10s of kpc.}} {we consider that the gas and CRs at most points in our simulation would be approximately co-spatial throughout most of the system's evolution. Thus, the number of secondary CRs produced for a quantity of gas and CRs in the evolved simulation would be reasonably approximated by a uniform interaction rate per CR particle. This allows us to adopt a uniform CR interaction rate over the bubble lifetime and hence estimate a representative volumetric injection rate of secondary electrons.}

{We therefore consider that 
the evolution of the secondary electrons injected by the pp interaction can be reasonably approximated by a steady-state scenario,} {where the injection of secondary electrons is practically constant and is matched by their cooling, i.e: }
\begin{equation}
{\frac{{\rm d}\gamma_{\rm e}}{{\rm d}t}\;\!\frac{\partial n_{\rm e}}{\partial \gamma_{\rm e}} + n_{\rm e} \frac{\partial}{\partial \gamma_{\rm e}}\left(\frac{{\rm d}\gamma_{\rm e}}{{\rm d}t}\right) = \Gamma_{\rm e} - \Lambda_{\rm e}} \ .
\label{eq:de_leptons}
\end{equation} 
{Here, $\Lambda_{\rm e}$ is introduced as the particle destruction term (this is set to $\Lambda_{\rm e} = 0$ in the case of CR electrons, as absorption processes are not significant), and the term ${\rm d}\gamma_{\rm e}/{\rm d}t$ specifies the energy loss (or cooling rate) of the particles,
being the sum of the cooling processes affecting the secondary electrons (these would be the same as for the primary electrons considered in the leptonic model -- adiabatic cooling, synchrotron, inverse Compton and bremsstrahlung).  
The electron injection rate is given by: }
\begin{equation}
{\Gamma_{\rm e}(\gamma_{\rm e}) =  \int_{\gamma_{\rm 0, p}}^{\gamma_{\rm 1, p}}\frac{\partial f(\gamma_{\rm e};\gamma_{\rm p})}{\partial \gamma_{\rm p}} \;\! \dot{n}_{\rm p\pi}(\gamma_{\rm p}) \;\! {\rm d}\gamma_{\rm p} \ ,}
\label{eq:injection_secondaries}
\end{equation}
{where the relative production fraction of electrons in the proton's rest frame for the pp interaction is computed using the publicly available code {\tt aafragpy}~\citep{Koldobskiy2021}.\footnote{{This code is based on {\tt Aafrag}~\citep{Kachelriess2019CoPhC}, but provides an extension to lower CR proton energies, below 4 GeV, using production parameterizations obtained by~\cite{Kamae2006ApJ, Kamae2007ApJ}.}}
By expressing equation~\ref{eq:de_leptons} as an integral over energy, we may 
write the steady-state spectral density of the secondary electrons as:} 
\begin{equation}
    \label{eq:electron_bc}
    {n_{\rm e}(\gamma_{\rm e}) = t_{\rm cool}(\gamma_{\rm e}) \int_{\gamma_{\rm e}}^{\gamma_{\rm e}^{\rm max}} {\rm d}\gamma_{\rm e}'\; \Gamma_{\rm e}(\gamma_{\rm e}') \ ,}
\end{equation}
{where $t_{\rm cool}$ is the total effective electron cooling timescale, and where we set $\gamma_{\rm e}^{\rm max}$ as the maximum secondary electron energy, as informed by the secondary production spectrum. We note that similar approaches in modeling secondary CR electron spectra have been used in other studies~\citep[e.g.][]{Wiener2013MNRAS}.}

The radiation emitted from galaxy bubbles in the hadronic scenario is comprised of two components: (1) the multiwavelength emission from the secondary electrons, and (2) $\gamma$-ray emission from the decay of neutral pions. In the first case, once the steady-state secondary CR electron spectrum is known from solving equation~\ref{eq:de_leptons}, the resulting emission spectrum is computed in the same way as for the CR electrons in the leptonic model. In the second case,
 $\gamma$-ray emission is caused by the decay of neutral pions. This proceeds (with a branching ratio of 98.8\%) as
  $\pi^0 \rightarrow 2\gamma$
   on a timescale of $8.5 \times 10^{-17}\;\!{\rm s}$ \citep{Tanabashi2018PRD} and, given the limited energy-dependence of the inclusive $\pi^0$ formation cross-section~\citep[see, e.g.][]{Kafexhiu2014}, yields a $\gamma$-ray spectrum closely following the shape of the underlying CR proton spectrum. 
 The differential $\gamma$-ray inclusive cross section for the $\gamma$-ray emission can be written as: 
   \begin{equation}
       \frac{{\rm d}{\sigma}_{\rm p\pi}(\gamma_{\rm p}, \epsilon)}{{\rm d}\epsilon} = \mathcal{P}(\gamma_{\rm p})\times\mathcal{F}(\gamma_{\rm p}, \epsilon)  \ ,
       \label{eq:diff_xs}
   \end{equation}
   where the peak function $\mathcal{P}$ and spectral shape function $\mathcal{F}$ are well-parametrised, to an accuracy of better than 10 per cent, by~\citet{Kafexhiu2014}.
  The spectral emissivity of $\gamma$-rays then follows as:
\begin{equation}
j_{\pi\gamma}(\epsilon) = 
\epsilon\;\! \lambda_{\rm c} \;\!m_{\rm e} c^2 \;\!n_{\rm H} \;\! \int_{\gamma_{\rm 0, p}}^{\gamma_{\rm 1, p}} \frac{{\rm d}{\sigma}_{\rm p\pi}(\gamma_{\rm p}, \epsilon)}{{\rm d}\epsilon}\;\! n_{\rm p}(\gamma_{\rm p}) \;\! {\rm d}\gamma_{\rm p} \ ,
\end{equation}
where $\lambda_{\rm c}$ is the Compton wavelength of an electron, $n_{\rm H}$ is the local density of thermal gas, and other terms retain their earlier definitions.\footnote{In-line with the earlier notation conventions, the appearance of the electron rest mass here results from the use of $\epsilon$ as the dimensionless photon energy in units of electron rest mass.}

\subsubsection{Hybrid model}
\label{sec:hybrid_bubbles}

In our lepto-hadronic hybrid composition model, we mix the primary CR composition to include both electrons and protons. The interactions and emission properties of the CR electrons and CR protons in this configuration are the same as in their respective pure-composition models introduced in sections~\ref{sec:leptonic_bubbles} and~\ref{sec:hadronic_bubbles}, however the normalization of the electron and proton spectra are now modified.  In the hybrid model, equation~\ref{eq:electron_norm} is replaced by
\begin{equation}
\mathcal{N}_e^{\rm hy} = f_{\rm lep} \;\! \mathcal{N}_e^{\rm lep}\ ,
\label{eq:electron_norm_hyb}
\end{equation}
where the scaling parameter $f_{\rm lep}$ is introduced to set the fraction of the CR energy density  $e_{\rm CR}$ contributed by CR electrons. As with the leptonic model, the electrons are aged to yield an appropriate spectrum for the simulated bubble age. Similarly, 
equation~\ref{eq:proton_norm} is replaced by
\begin{equation}
\mathcal{N}_p^{\rm hy} = f_{\rm had} \;\! \mathcal{N}_p^{\rm had}
\label{eq:proton_norm_hyb}
\end{equation}
where the scaling parameter $f_{\rm had}$ sets the fraction of CR energy density contributed by CR protons. Other terms in equations~\ref{eq:electron_norm_hyb} and~\ref{eq:proton_norm_hyb} retain their earlier definitions. 
It follows that, for a self consistent hybrid model, $f_{\rm lep} + f_{\rm had} = 1$. 
The resulting emission from both the hadronic and leptonic components are then combined to give the hybrid model spectrum.

\section{Results}
\label{sec:section3}

\subsection{Physical characteristics of the simulated bubbles}
\label{sec:bubble_development}

Our simulation results for the bubble evolution and their MHD properties are shown in~\citetalias{OwenYang2021MNRAS} (in particular, their Figure 1 and 2, i.e. `Run A'), to which we refer the reader for detailed discussion. 
Here, we provide a summary, emphasizing aspects of the MHD properties of the bubbles and their evolutionary progression of particular consequence to the leptonic and hadronic emission discussed later in sections~\ref{sec:spectra} and~\ref{sec:broadband_emission_maps}. 

The initial expansion of the bubbles is driven by an injection of CR energy, thermal gas and magnetic field energy provided by bipolar jets at the center of the simulation domain. These are directed in the $\pm z$ directions, perpendicular to the plane of the host galaxy, and remain active for 0.3 Myr (cf. the simulation set up of~\citetalias{Yang2012ApJ};~\citealt{Guo2012ApJ}, and the magnetic field sub-grid model of~\citealt{Sutter2012MNRAS}).
The resulting bubbles are over-pressured with respect to their ambient medium. {Their expansion continues after the jets have shut down and is supersonic,} reaching Mach numbers of around $M\sim 30$ by 1 Myr in the vertical direction. The lateral expansion of the bubbles is slower, but is also supersonic ($M\sim 10-12$). The resulting forward shocks at the bubble surfaces compresses hot ambient halo gas into a shell around the bubbles, of temperature $T>10^8$ K, and densities $10^{-2}\;\!{\rm cm}^{-3}$. 
A magnetic draping layer~\citep{Lyutikov2006MNRAS} forms at the shocked shell, where magnetic field lines are stretched and compressed, strengthening and aligning the field parallel to the expanding shell. The fast-flowing gas within the bubbles advects CR energy density to their leading upper surface. These CRs diffuse along local magnetic field vectors and form a sharp `edge' in CR energy density at the leading expansion front, where the draping layer practically contains CRs within the bubbles. 
Behind the shocked shell, an outer contact discontinuity encloses 
lower temperature $T\geq 10^7\;\!{\rm K}$, $n\sim 10^{-4}\;\!{\rm cm}^{-3}$ gas within the bubbles.  
At the center of the bubbles, large under-dense lobes form. These lobes are surrounded by an inner contact discontinuity, and contain very high temperature $T>10^8$ K jet plasma and jet-entrained halo gas. The lobes expand with the bubble and cool slowly, with their high temperatures persisting for the duration of the simulation, to 7 Myr.

\begin{figure*}
\includegraphics[width=0.9\textwidth]{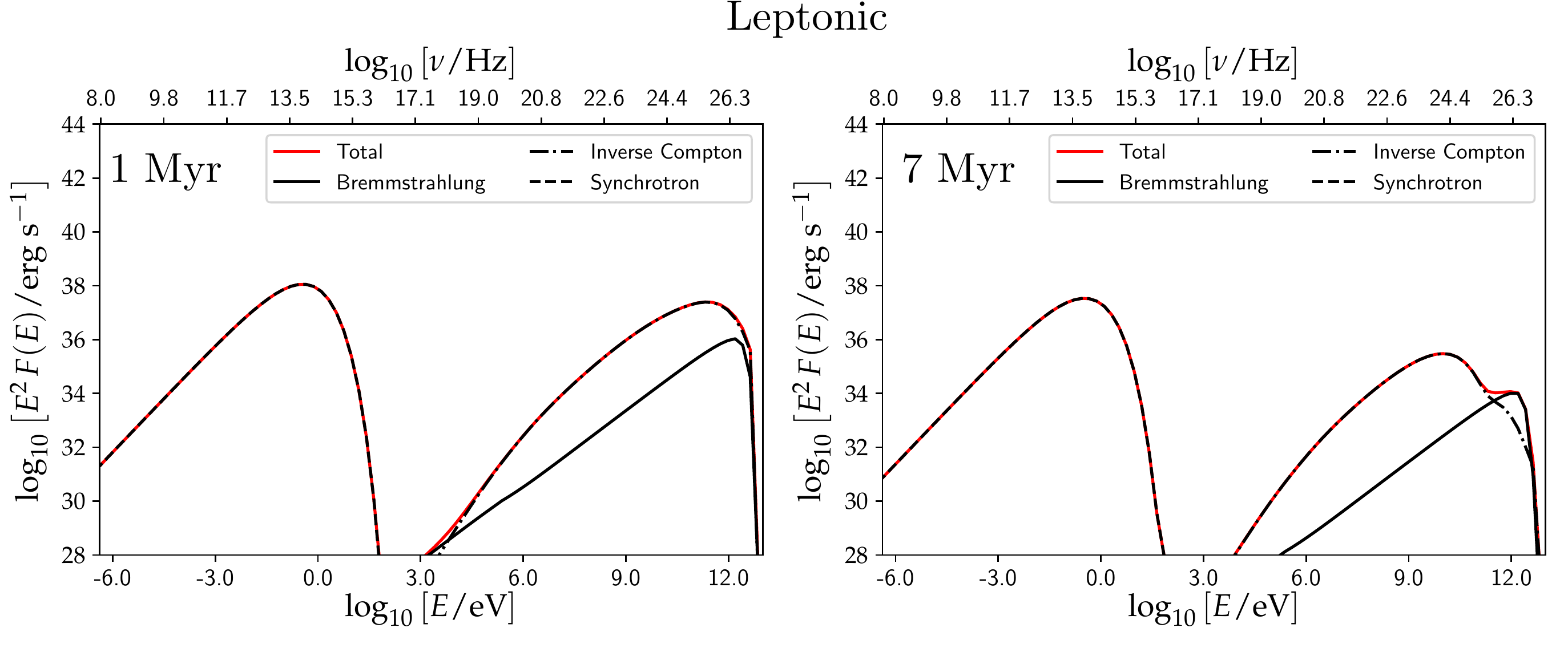}
\caption{{Emission spectrum from a leptonic bubble, showing contributions from {non-thermal} bremsstrahlung, synchrotron and inverse Compton. The left panel shows the expected emission spectrum at 1 Myr. The right panel shows the emission spectrum after 7 Myr. The top abscissa shows the emitted photon frequency, and the bottom abscissa shows the corresponding photon energy.}}
\label{fig:fig_lepton_1_7Myr}
\end{figure*}

\subsection{Bubble spectra}
\label{sec:spectra}

We post-process our MHD bubble simulations to compute their expected emission properties in 1 Myr intervals, but put focus on the differences in their emission properties during their early (1 Myr) and late (7 Myr) stages of evolution. While we relate our computed emission results to the physical properties of the simulated bubbles, we pay particular attention in this paper to the  
comparison between the emission properties of the bubbles under leptonic, hadronic and hybrid lepto-hadronic composition scenarios.

We compute the volume emissivity at each point throughout our simulation grid 
under three different CR composition scenarios: purely leptonic ($f_{\rm lep} = 1$, $f_{\rm had} = 0$), purely hadronic ($f_{\rm lep} = 0$, $f_{\rm had} = 1$), and a hybrid lepto-hadronic mix ($f_{\rm lep} = 0.5$, $f_{\rm had} = 0.5$). In the hybrid case, our choice of an equal proportion of CR energy density in hadrons and leptons is intended to reflect a mid-point of possible model configurations. With no clear constraint on the balance between CR species in external galaxy bubbles, we consider that 
a different choice would not be any more or less physical. Moreover, emission properties of hybrid bubbles with a different ratio of hadrons to leptons can be reasonably estimated by weighting the emission components we present here. 
In each case, we include all processes described in section~\ref{sec:cr_spec_int} as appropriate (synchrotron, inverse Compton and non-thermal bremsstrahlung and, in the case of hadronic or hybrid compositions, pion-decay $\gamma$-rays). 
We integrate over the full simulation volume to capture the emitted spectrum from the CRs in our bubbles. As the CR energy density is well confined to the bubble region, the emission volume of the simulation is completely dominated by the bubbles themselves. 
We find no significant differences in our results if we instead constrain the integration volume to match the extent of the bubbles. 

\subsubsection{Leptonic bubble}

Figure~\ref{fig:fig_lepton_1_7Myr} shows the emitted spectrum from a leptonic bubble at 1 Myr and 7 Myr. 
The spectrum at lower energies is dominated by synchrotron emission, which is persistent through to 7 Myr. There is a synchrotron peak at optical wavelengths, however this would likely be undetectable around external galaxies, being relatively diffuse and substantially less luminous than their stellar emission. The synchrotron emission extends to microwave and radio frequencies, where detection prospects would be substantially better (see, e.g.~\citetalias{OwenYang2021MNRAS}, which found that radio emission from bubbles of the same configuration could be detected out to distances of $\sim$20 Mpc with up-coming facilities like the Square Kilometer Array, SKA). 
Our model at 1 Myr (a similar age to the Galactic  \textit{Fermi} bubbles, as suggested by e.g.~\citetalias{Yang2012ApJ}) predicts a radio power spectral density of $\sim 10^{16}$ W Hz$^{-1}$ at 408 MHz, which is lower than that estimated for the \textit{Fermi} bubbles (see, e.g.~\citealt{Haslam1982AAS, Jones2012ApJl}). Moreover, the microwave haze associated with the Galactic \textit{Fermi} bubbles also exceeds the emission predicted here. 
At 23 GHz, for example, latitude-dependent emission of between $\sim$0.2-6 kJy sr$^{-1}$ was reported by \textit{WMAP}~\citep{Dobler2008ApJ}. If this were attributed to a soft synchrotron origin, it would exceed our model prediction by a similar factor to the 408 MHz radio emission. As the synchrotron emission from a bubble is regulated by its magnetic field strength, this may indicate the magnetic field of the Galactic \textit{Fermi} bubbles are stronger than in our model. Indeed, previously, in~\citetalias{Yang2013MNRAS}, it was shown that the microwave haze emission of the \textit{Fermi} bubbles could be matched when the default {\tt GALPROP}~\citep{Strong2007ARNPS} exponential model for the magnetic field was used instead. We thus consider that our current model may be regarded as a conservative lower limit of the synchrotron emission in microwave and radio bands. 

At intermediate energies, Figure~\ref{fig:fig_lepton_1_7Myr} shows {that non-thermal} bremsstrahlung processes become more important. 
We find this dominates the keV X-ray emission during the first $\sim 2$ Myr. {While \citetalias{OwenYang2021MNRAS} demonstrated this non-thermal emission would likely not be detectable in external galaxies, X-ray observations towards the Galactic \textit{Fermi} bubbles reveal substantially higher emission than that computed with our model at a similar bubble age (see the comparison in~\citetalias{OwenYang2021MNRAS};  also~\citealt{Snowden1997ApJ, Kataoka2013ApJ, Predehl2020Natur}). Although this may suggest better X-ray detection prospects for external galaxy bubbles than our results would imply, we note that these Galactic observations may include a very significant thermal bremsstrahlung contribution from all the gas in the Milky Way halo, which likely extends to a radius of $\sim 250\;\!{\rm kpc}$~\citep[i.e. far larger than the size of our simulation box, see e.g.][]{Blitz2000ApJ, Grcevich2009ApJ}, the Galactic bulge and gas heated by the shocks associated with the bubble (\citealt{Yang2022NatAs}; see also~\citealt[][]{Zhang2021ApJ}), none of which is included in our model (and could evolve as the bubble ages). 
Moreover, features external to the Galactic bubbles (e.g. the North Polar Spur;~\citealt{Kataoka2013ApJ}) are not included in our model, but may make a contribution to the observed X-ray emission. We thus consider a direct comparison between the X-ray emission from our model and that observed from the Galactic \textit{Fermi} bubbles/halo to be complicated by substantial thermal emission and emission from structures not associated with the bubbles. Addressing these additional contributions is non-trivial, and is not necessarily informative to predict the observational prospects of bubbles around external galaxies where thermal emission contributions could differ greatly. } 
After the first few Myr, {non-thermal} bremsstrahlung X-rays are surpassed by inverse Compton emission, which dominates the X-ray emission from the bubble by 7 Myr. 
At these later times, a significant non-thermal bremsstrahlung component can {also} be seen to emerge in TeV $\gamma$-rays. This is attributed to the concentration of gas and CR energy density near the top of the bubble (cf. section~\ref{sec:broadband_emission_maps}), although we note that the physical strength of this emission may not be properly resolved.\footnote{For example, a higher resolution simulation may resolve differences between the gas and CR energy density distribution, which a lower resolution realization would attribute to the same cell.}
This will be explored in more detail in future work (see also section~\ref{sec:remarks}). 

The $\gamma$-ray emission from a leptonic bubble throughout its lifetime is driven by inverse Compton scattering (with the exception of the possible TeV non-thermal bremsstrahlung contribution at late times). 
At energies above a few {10s of} GeV, the ISRF of the host galaxy is the primary source of target photons. This is spatially-dependent, and thus introduces significant evolutionary behavior into the resulting inverse Compton spectrum. As the bubble expands, it moves much of its entrained CR energy density to higher altitudes where the ISRF is weaker. This causes the associated inverse Compton emission to drop rapidly (cf. the high energy emission at 1 Myr and 7 Myr in Figure~\ref{fig:fig_lepton_1_7Myr}). 
At {lower $\gamma$-ray energies}, the inverse Compton emission is instead attributed to up-scattered CMB photons. As the CMB is spatially homogeneous, it does not imprint any spatial variation into this emission. Thus, any spectral evolution in this regime must be a consequence of (adiabatic) cooling of the CR electrons.\footnote{This was included in our post-processing model, as outlined in section~\ref{sec:leptonic_bubbles}; for full details, see~\citetalias{OwenYang2021MNRAS}.} 
{At 1 Myr, the overall emission in the $\gamma$-ray band, integrated over the bubble between 1 and 100 GeV, is $\sim 10^{37} {\rm erg}\;\!{\rm s}^{-1}$, which is in good agreement with the $\gamma$-ray emission observed from the Milky Way's \textit{Fermi} bubbles over a similar energy range~\citep{Su2010ApJ}.}

\begin{figure*}
\includegraphics[width=0.9\textwidth]{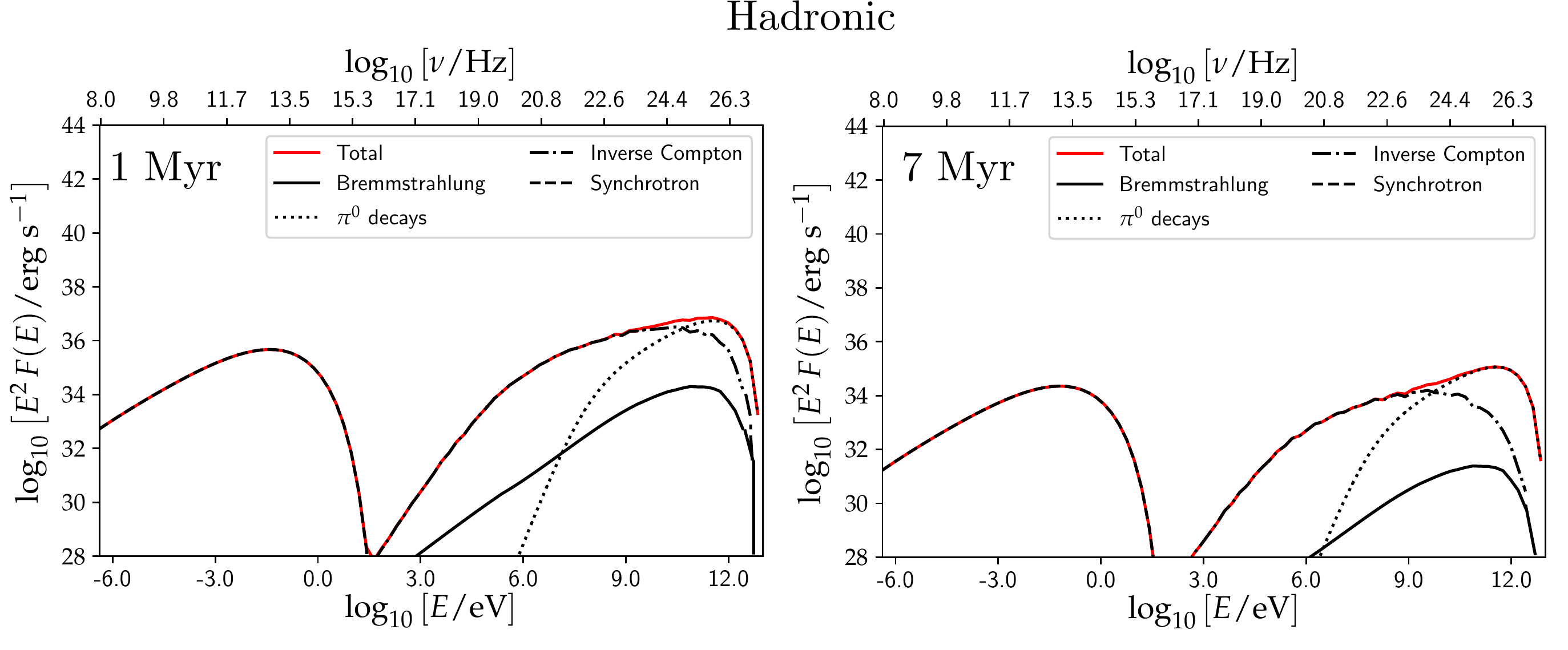}
\caption{{As figure~\ref{fig:fig_lepton_1_7Myr}, but for a hadronic bubble. This shows the contributions from bremsstrahlung, synchrotron and inverse Compton emission, mediated by secondary CR electrons produced in hadronic interactions (via the decay of charged pions, $\pi^{\pm}$). An additional component, arising from the decay of neutral pions $\pi^0$ to $\gamma$-rays is also shown (as labeled).}}
\label{fig:fig_hadron_1_7Myr}
\end{figure*}

\begin{figure*}
\includegraphics[width=0.9\textwidth]{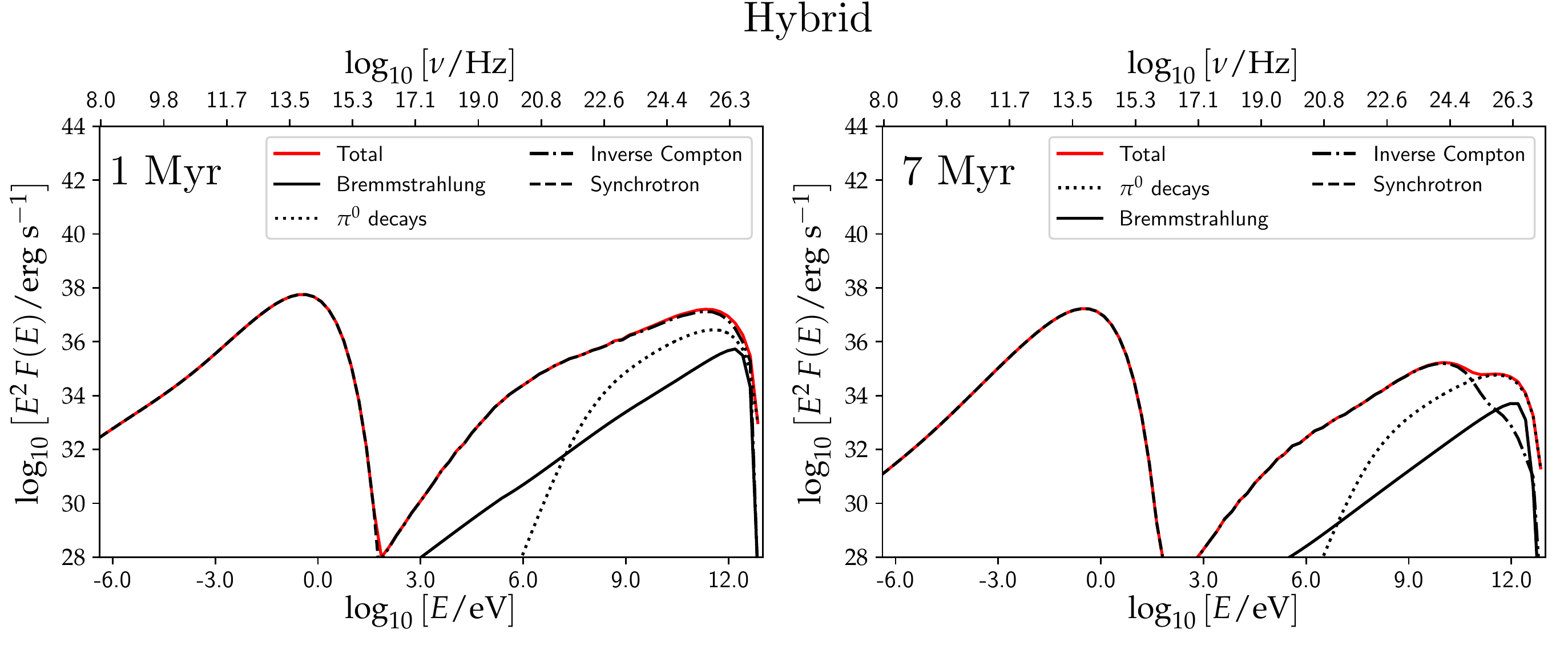}
\caption{{As figure~\ref{fig:fig_lepton_1_7Myr}, but for a hybrid lepto-hadronic bubble, adopting an equal CR composition of protons and electrons, i.e. $f_{\rm lep} = f_{\rm had} = 0.5$. The spectral components inherited from the hadronic and leptonic bubbles, as shown in Figures~\ref{fig:fig_lepton_1_7Myr} and~\ref{fig:fig_hadron_1_7Myr}, are evident (inverse Compton, synchrotron and bremsstrahlung from both primary and secondary electrons, and pion $\pi^0$ decay $\gamma$-rays from hadronic interactions).}}
\label{fig:fig_hybrid_1_7Myr}
\end{figure*}

\subsubsection{Hadronic bubble}
\label{sec:hadronic_bubble_spec}

{Figure~\ref{fig:fig_hadron_1_7Myr} shows the spectrum for a hadronic bubble. This differs noticeably from its leptonic counterpart in two respects.} {Firstly, the synchrotron and inverse Compton peaks arise at lower energies.} This reflects a lower spectral cut-off in the freshly injected secondary electrons of the hadronic model. The electron spectrum in the leptonic model is aged with the bubble, while secondary electrons in the hadronic scenario are injected assuming a steady-state at local conditions. The favored injection locations of these secondaries are typically where densities (and CR interaction rates) are higher. In~\citetalias{OwenYang2021MNRAS} (their Figure 1), it can be seen that regions of higher density in the bubble tend also to harbor stronger magnetic fields. These drive higher synchrotron losses than provided by the `averaged' ageing approach applied to the electron spectrum in the leptonic model, thus returning a slightly {reduced upper cut-off} and steeper electron spectrum.  

{Secondly}, a pion-decay bump is present, and this contributes significant (by 7 Myr, dominant) $\gamma$-ray emission in the GeV-TeV energy range. This component ensures that the $\gamma$-ray emission between $\sim$10 GeV and $\sim$1 TeV fades much more slowly than in a leptonic bubble. Pion-decay $\gamma$-rays are a close tracer of the hadronic interactions~\citep[e.g.][]{Owen2021MNRAS}. They rely on engagement between CR protons and ambient gases. Thus, such emission will be strongest from regions where CR energy density and gas density is high. As our simulations advect CR energy density with the thermal gas, any divergence between these quantities is driven solely by CR diffusion. CR diffusion timescales are much longer than advection timescales inside the bubble, hence pion-decay emission will remain comparatively stable over the lifetime of the bubble. Indeed, this can be seen in Figure~\ref{fig:fig_hadron_1_7Myr} at 7 Myr (right panel), where the pion-decay component remains relatively strong, and has reduced far less, by a factor of $\sim 10^2$, than TeV $\gamma$-rays contributed by inverse Compton scattering, which fall by a factor of {$\sim 10^3$} compared to the 1 Myr spectrum (left panel).

\subsubsection{Hybrid bubble}

Figure~\ref{fig:fig_hybrid_1_7Myr} shows the 1 and 7 Myr spectra for the hybrid composition model. This inherits the pion-decay emission from the hadronic model, which dominates the late-time $\gamma$-ray emission between {a few 10s of} GeV and $\sim 1$ TeV. At 1 Myr, the inverse Compton component and pion decay emission at GeV energies are comparable, but would show distinguishable spectral differences if more sophisticated ISRF models were adopted. Such differences have been considered in other works, which put focus on the Galactic \textit{Fermi} bubbles~\citep[e.g.][]{Ackermann2014ApJ, Lunardini2015PhRvD,Yang2017ApJ, Abeysekara2017ApJ} and invoked more detailed ISRF models appropriate for the Milky Way~\citep[e.g., that used in {\tt GALPROP};][]{Strong2007ARNPS} to show that a purely hadronic scenario would be difficult to reconcile with the observed $\textit{Fermi}$-LAT spectrum of the Galactic bubbles. 

\subsection{Emission maps}
\label{sec:broadband_emission_maps}

\begin{figure*}
\includegraphics[width=0.75\textwidth]{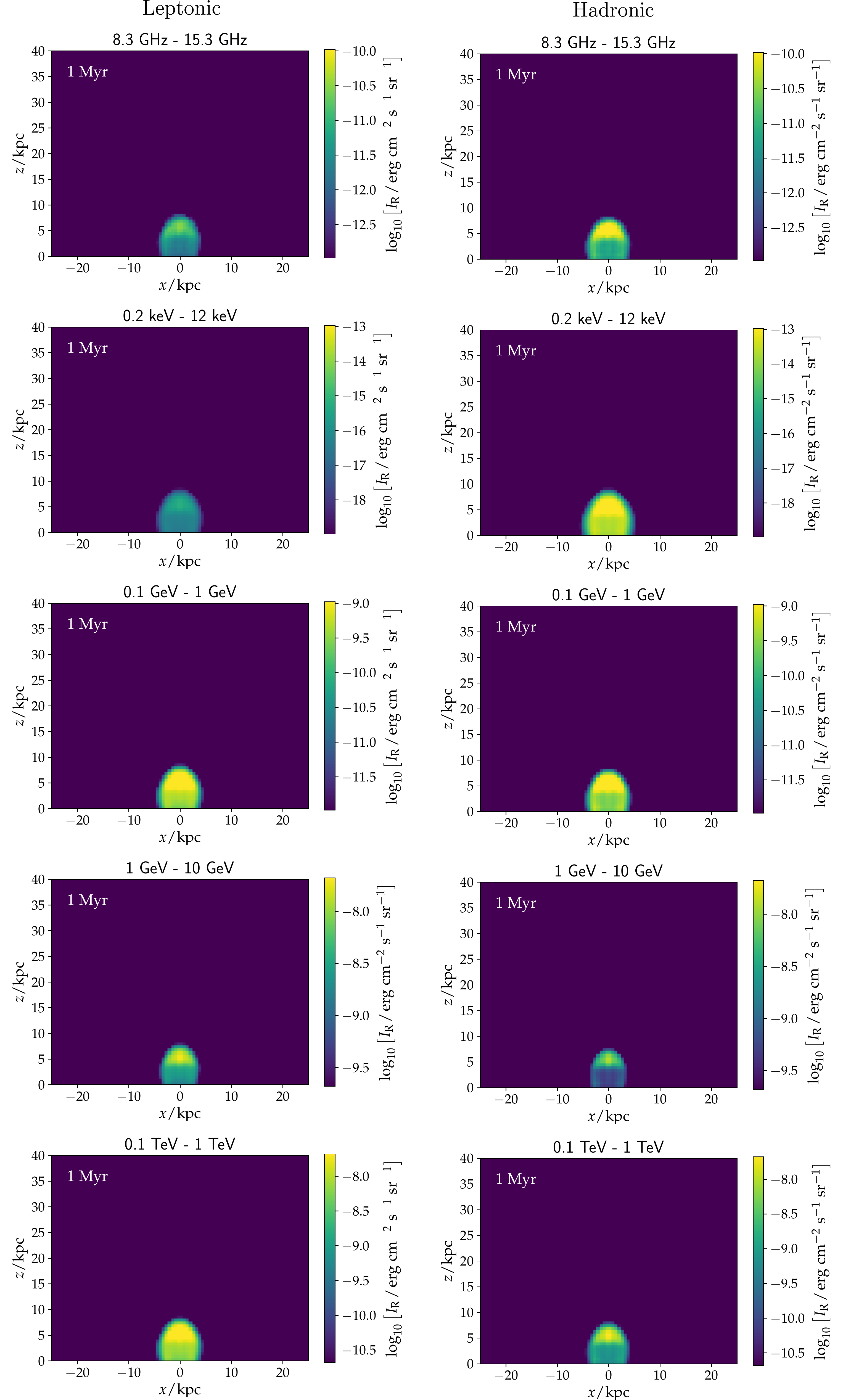}
\caption{{Broadband emission maps shown as 2D projection plots at 1 Myr. Panels on the left show the case for the leptonic model, while those on the right are for the hadronic model. The energy bands are indicated for each plot. To aid morphological comparison with earlier works, this assumes that the bubble is located at a distance of 8 kpc (as in~\citetalias{Yang2012ApJ}), however projection effects are not considered.}}
\label{fig:1Myr_leptonic_hadronic_compare}
\end{figure*}

\begin{figure*}
\includegraphics[width=0.75\textwidth]{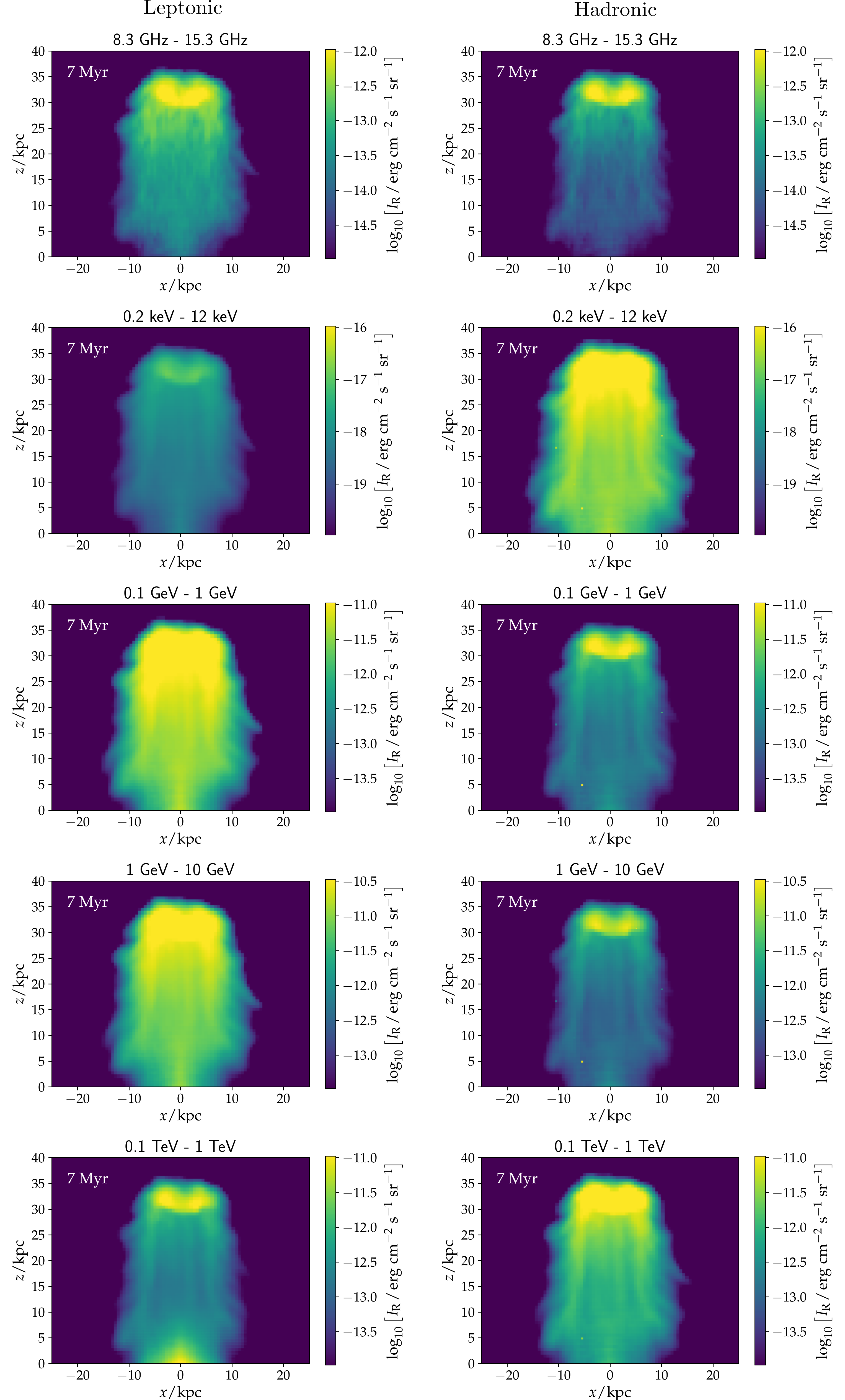}
\caption{{Same as Figure~\ref{fig:1Myr_leptonic_hadronic_compare}, but instead showing the bubble emission structure at 7 Myr.}}
\label{fig:7Myr_leptonic_hadronic_compare}
\end{figure*}

We compute spatial emission maps for the simulated bubbles from the volume emissivity throughout the simulation grid under leptonic and hadronic emission scenarios. The hybrid case is not explicitly considered here, as its emission structure can be inferred from the combination of the purely hadronic/leptonic cases. We integrate the volume emissivity throughout the simulation grid over one spatial dimension (thus producing a 2D projected emission map), and over a specified energy range. We consider five energy bands, covering radio frequencies (Band A: 8.3 - 15.3 GHz), X-rays (Band B: 0.2 - 12 keV) and $\gamma$-rays {(Band C: 0.1-1 GeV, Band D: 1-10 GeV and Band E: 0.1 - 1 TeV)}. These are chosen to capture the emission structure arising from the different physical processes underlying the bubble emission, but also reflects the energy bands that will be accessible with current and up-coming observational facilities, namely SKA (radio), \textit{Athena} (X-rays), \textit{Fermi}-LAT in $\gamma$-rays {(to $\sim$100 GeV)}, and the Cherenkov Telescope Array (CTA) for higher energy $\gamma$-rays.

\subsubsection{Early-stage evolution}
\label{sec:early_state_map}

Figure~\ref{fig:1Myr_leptonic_hadronic_compare} compares the spatial emission of a hadronic and leptonic bubble at 1 Myr in each of the five energy bands. This shows that many aspects of the emission structure are common to both the hadronic and leptonic models. For example, the radio 8.3 - {15.3} GHz emission is brighter at the top and surface of the bubble, where the combination of the magnetic field and CR energy density is greatest. The X-ray and {1-10 GeV} $\gamma$-ray emission is also {particularly} concentrated towards the top of the bubbles, but shows somewhat more {varied} distribution throughout the bubble volume compared to the radio emission. The X-ray emission is driven by {non-thermal} bremsstrahlung and a non-negligible inverse Compton component formed from up-scattered CMB photons {(especially in the hadronic model)}, while the {1-10 GeV} $\gamma$-ray band is dominated by the latter of these. The CMB photons, being homogeneous, do not introduce any spatial dependence to the bubble emission structure, thus the  GeV $\gamma$-ray band primarily traces the CR energy density through the bubble. The X-ray emission distribution differs slightly, in that the {non-thermal} bremsstrahlung emission is also weighted by gas density. {We note that thermal bremsstrahlung emission from shocks would also be expected in the X-ray band. However, as this thermal component is not included in our calculations (and we do not invoke CR acceleration at the bubble shocks), the shocks are not visible in our X-ray emission maps.}

Figure~\ref{fig:1Myr_leptonic_hadronic_compare} also reveals certain differences between the hadronic and leptonic bubbles. Firstly, the emission is generally brighter for the leptonic bubble {in higher energy bands, but dimmer in lower energy bands} (as is also {reflected by} their spectra, see section~\ref{sec:hadronic_bubble_spec}). {Secondly,} the hadronic bubble typically presents more spatial variation in {the intensity of its emission} which more closely reflects the underlying hydrodynamical structure. 
Both of these differences can be attributed to the origin of the emission. In the leptonic case, emission is driven by primary CR electrons, which trace the CR energy density directly. In the hadronic case, the emission is instead driven by the provision of secondary CR electrons by hadronic interactions. This is set by the local density and CR energy density, thus hadronic emission will be stronger in regions where both CR energy density and gas density is higher. As such, the hadronic emission map more clearly reflects the hydrodynamical structure of the bubble, with the top and surface layers being more visually defined, and with the contrast between the surface emission and internal bubble cavity emerging more clearly. This difference is particularly evident in the {0.1-1 TeV} band, where it is further highlighted by the presence of pion decay emission. This makes a significant contribution to the {0.1-1 TeV} $\gamma$-ray emission in the hadronic model, but is not present at all in the leptonic case. 

\subsubsection{Late-stage evolution}

The hadronic and leptonic emission maps for the bubble at 7 Myr are shown in Figure~\ref{fig:7Myr_leptonic_hadronic_compare}. Compared to the 1 Myr emission in Figure~\ref{fig:1Myr_leptonic_hadronic_compare}, the older bubbles are substantially larger, and show dimmer yet more structurally-diverse emission. Morphological features of the bubbles can be seen in more detail than at 1 Myr, and more variation is revealed between wavebands, as well as between the two composition scenarios. 

{The most striking differences between the leptonic and hadronic bubble emission can be seen in the highest energy 0.1-1 TeV $\gamma$-ray band, where the 7 Myr hadronic emission is brighter than its leptonic counterpart. This is driven by pion decays, which preferentially trace the top of the bubble, where both CR energy density and gas density are particularly high. At low altitudes, a small amount of inverse Compton emission is also present, but this remains sub-dominant. In the leptonic bubble, the emission separates into two spatial components. Much of this is contributed by non-thermal bremsstrahlung at the top of the bubble, which dominates in the absence of pion decays. The emission at the base of the bubble is due to inverse Compton scattering in the ISRF, which is barely seen in the hadronic model (although present, it is very sub-dominant). In lower $\gamma$-ray energy bands (0.1-1 GeV and 1-10 GeV), the leptonic bubble is brighter and shows less spatial variation than the hadronic case. This follows from the distribution of secondary electron injection in the hadronic model, which boosts emission from higher density regions. The brightness of the bubbles is reversed at X-ray energies, with inverse Compton emission remaining important in the hadronic model (see also Figure~\ref{fig:fig_hadron_1_7Myr}), but being broadly absent from this band in the leptonic case. } In all energy bands, there is a strong peak of intensity at the top of the bubble, where the highest CR energy densities, gas densities and magnetic field strengths {coincide}. These conditions enhance emission from bremsstrahlung, inverse Compton and (in the case of stronger magnetic fields) synchrotron processes. In the hadronic model, these higher densities also promote the injection of secondary electrons. 
The hadronic model is therefore more sensitive to the underlying gas density (as was also see in the early-stage evolution in Figure~\ref{fig:1Myr_leptonic_hadronic_compare}), with both models also reflecting the distribution of CR energy density. 

The 7 Myr radio emission is perhaps the most morphologically complex. However, the structures evident in Figure~\ref{fig:7Myr_leptonic_hadronic_compare} are common to both the leptonic and hadronic model. At these frequencies, the emission remains 
synchrotron-dominated, and is driven by CR electrons cooling in magnetic fields. 
Although the excess at high altitudes remains relatively strong due to the persistent high CR energy density at the bubble leading edge, much more sub-structure is discernible throughout the bubble. In particular, the edge of the emission region shows a feathered morphology and, across the emission surface, projected filamentary structures can be seen aligned vertically in the prevailing flow direction of the internal gas.
Both of these features were found to be a consequence of the magnetic field structure (see~\citetalias{OwenYang2021MNRAS} for discussion), particularly at the contact discontinuity within the bubble which practically marks the edge of the main emission zone. In this region, the magnetic field is relatively disordered, which leads to uneven CR diffusion near the contact discontinuity. This creates the feather-like emission surface seen in Figure~\ref{fig:7Myr_leptonic_hadronic_compare}, with brighter filaments emerging within these features, following the structure of the magnetic field. The vertical filaments within the emission region are a consequence of the same effect. These are the face-on feathered emission filaments located mainly on the bubble's outer contact discontinuity, which are only present when the coherence length of the surrounding halo magnetic field is small {compared to the size of the bubble} (\citetalias{OwenYang2021MNRAS}).  

\section{Discussion}
\label{sec:section4}

\subsection{CR energy spectrum}

\subsubsection{Spectral cut-off}
\label{sec:upper_cut_off}

 \begin{figure*}
\includegraphics[width=0.9\textwidth]{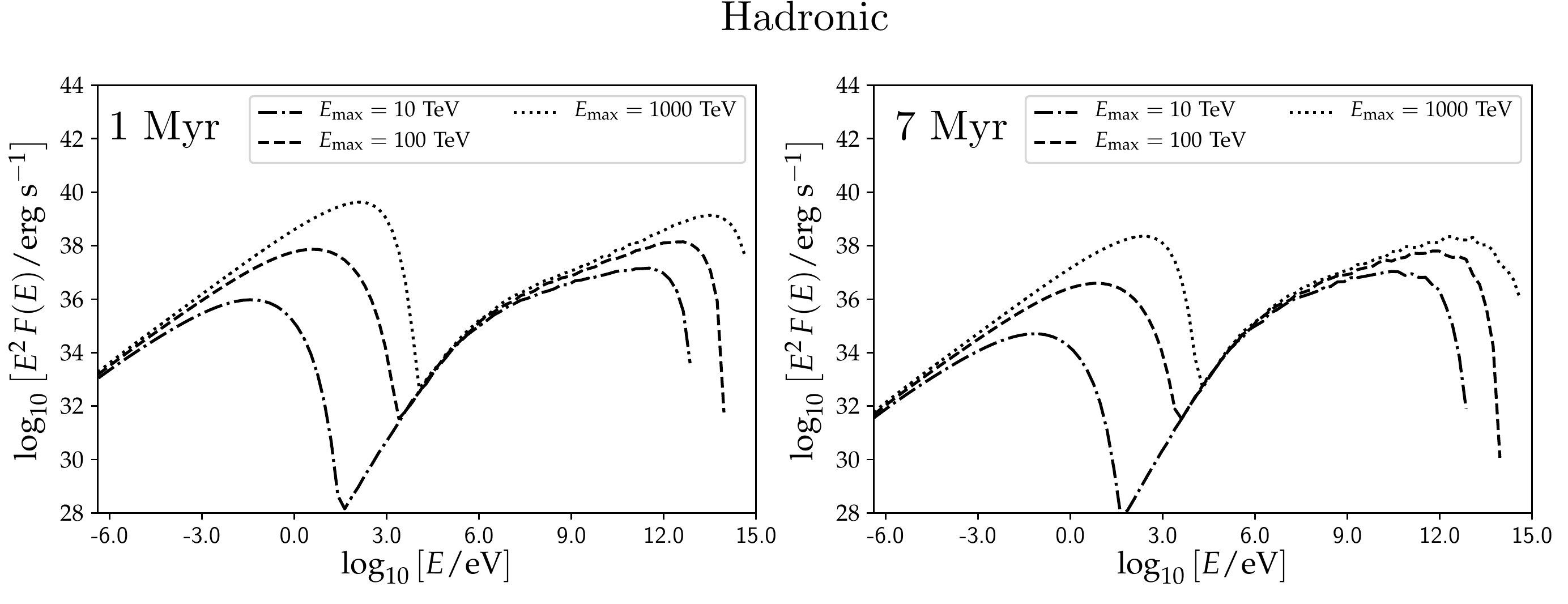}
\caption{{Hadronic bubble model at 1 Myr (left) and 7 Myr (right), adopting different maximum energies ($E_{\rm max} = 100$ and 1,000 TeV) for the proton spectrum, compared to the fiducial model  ($E_{\rm max} = 10$ TeV, as shown in Figure~\ref{fig:fig_hadron_1_7Myr}). Lines are shown up to a maximum resolvable energy, beyond which the spectrum could not be properly reconstructed with the adopted energy resolution of our post-processing calculation.}}
\label{fig:max_cutoff}
\end{figure*}

In both the leptonic and hadronic models, we 
adopted a power law spectrum for the CRs between 1 GeV and 10 TeV. The leptonic spectrum was used as an initial condition, which was then evolved to the required age of the bubble accounting for radiative cooling processes. This led to the natural emergence of a spectral cut-off at $\sim$ 1 TeV (see~\citetalias{OwenYang2021MNRAS}). 
Protons undergo radiative losses at a much lower rate than electrons~\citep[e.g.][]{Rybicki1979_book} and their spectral evolution throughout the bubble lifetime is negligible. Thus an upper limit to the proton spectrum would instead be inherited from their source environment and/or acceleration mechanism. 
This may be estimated by consideration of the physical origin of the CR protons in a hadronic bubble. Several possibilities have been discussed in the literature, in the context of accelerators for CR protons in hadronic models of the Galactic \textit{Fermi} bubbles. These mechanisms would presumably also apply in the external galaxy setting considered in this work, and fall into {three} categories: (1) maximum energies associated with an interstellar CR population supplied to the bubble; (2) maximum energies associated with acceleration processes operating alongside the bubble-inflating mechanism (in our case, the initial AGN jet){, and; (3) maximum energies associated with acceleration processes operating at the shock front.} 

The first of these categories does not invoke any particular assumptions about the bubbles themselves (at least in regards to the supply of CR protons), and instead considers that the origin of the CRs lies inside the host galaxy. These would then be advected out, into the bubble. Further re-acceleration may be possible within the advective flows or the  bubbles themselves~\citep{Cheng2012ApJ}, but only a small fraction of the CRs would be re-accelerated so the upper cut-off derived from galactic acceleration sources would remain applicable to the bulk of CR protons in the bubble. 
In this scenario, the value of $E_{\rm max}$ would  be set by likely interstellar accelerators, and these are mainly associated with stellar populations and/or their end-products. 
For example,~\citet{Cheng2012ApJ} considered a model where supernova remnants in the host galaxy accelerate CRs to $\sim \;\!{\rm PeV}$ energies. Lower or higher limits are also possible -- for instance, the maximum energy of CR protons attained in standard galactic supernova remnants has been demonstrated to reach $\sim 10-100$ TeV ~\citep{Lagage1983AA, Berezhko2000AA}. Alternatively,~\cite{Aharonian2019NatAs} proposed that a higher maximum energy of 1-10 PeV could be achieved by associations of massive star regions (based on recent $\gamma$-ray observations). Beyond this, evidence of hadronic PeV CRs within the Galaxy have also recently emerged, though their exact source (apart from the Crab nebula) remain uncertain (see \citealt{Cao2021Natur}; also~\citealt{AbdallaPoS2021} for a possible source candidate). CR acceleration in winds of star clusters has also been considered able to reach PeV energies in powerful cases~\citep{Morlino2021MNRAS}, while~\cite{Peretti2020MNRAS} argued that yet higher maximum energies, of 10s of PeV (and perhaps as high as 50-100 PeV), would be plausible in the nuclei of starburst galaxies due to the high level of turbulence expected in such regions. However, they also noted that acceleration up to these very high energies, beyond $\sim$ 10 TeV is non-trivial.

The second category of $E_{\rm max}$ constraints is derived from the mechanism used to initially inflate the bubble and supply CR energy density to the simulation, i.e. the intense explosive outburst from the center of the host galaxy. In our work, we model this to emulate bipolar AGN jets, active for 0.3 Myr, which could be regarded as the \textit{source} of CRs, rather than simply an agent to advect them into the bubbles.\footnote{{Other mechanisms to accelerate CRs in bubbles around galaxies have also been discussed. For example,~\cite{Romero2018AA} and~\cite{Muller2020MNRAS} considered that bubbles around some starburst galaxies could be formed by a superwind. The associated shocks and turbulent gas region of the resulting bubble might then accelerate CRs up to energies of $\sim 10^{18}\;\!{\rm eV}$ (under special circumstances).}} CRs supplied in this manner would therefore be disconnected from any constraints associated with possible CR accelerators within the host galaxy. Instead, the physical conditions within the jet would set $E_{\rm max}$. While theoretical or observational acceleration limits within young AGN jets are uncertain, some insight can be gained by scaling results from numerical simulations. For example, the 2D jet simulations of~\citet{Matthews2019MNRAS} find a maximum Hillas~\citep{Hillas1984ARAA} energy of $9.7\times 10^{18}\;\!{\rm eV}$ for a jet with velocity $0.95 c$, radius 1 kpc, power $3.89\times 10^{44}\;\!{\rm erg}\;\!{\rm s}^{-1}$ and characteristic magnetic field of 15.18 $\mu$G. The value of $E_{\rm max}$ associated with a jet then scales with the square root of the jet power~\citep{Matthews2019MNRAS}, its size~\citep{Bell2013MNRAS} and its characteristic magnetic field strength~\citep{Hillas1984ARAA}. Thus, from the~\cite{Matthews2019MNRAS} result, we can estimate that the jet model adopted in our simulations would impart a limit of 
\begin{equation}
E_{\rm max} \sim 4.7 \left( \frac{v_j}{0.025c}\right)\;\! \left( \frac{r_j}{0.5 \;\!{\rm kpc}} \right) \;\!\left( \frac{\langle|B|\rangle}{1\;\!\mu{\rm G}} \right) ~{\rm PeV} \ ,
\label{eq:emax_est}
\end{equation}
where the jet parameters used in our simulations are given in~\citet{Yang2012ApJ} and~\citet{Sutter2012MNRAS}.
This suggests a hadronic spectral cut-off at around 1 PeV (= 1,000 TeV) would be reasonable choice of $E_{\rm max}$, if CR protons were supplied in this way.

{The third category of $E_{\rm max}$ constraints is associated with CRs accelerated at the shock fronts
(e.g.~\citealt{Zhang2020ApJ}), particularly at the leading edge of the bubbles, where Mach numbers are highest (\citetalias{OwenYang2021MNRAS}) and CR acceleration would be most efficient~\citep{Zhang2020ApJ}. This process could operate to provide a source of freshly accelerated CRs over time, which would mainly be supplied to the outer reaches of the bubble at the location of the shock.\footnote{{Note that the shock front and bubble surface do not necessarily coincide; cf. the Galactic \textit{Fermi} and \textit{eRosita} bubbles (see e.g. \citealt{Yang2022NatAs}).}} This would lead to differences in the spatial emission patterns compared to those obtained in this work. In particular, 
outer edges of the bubble and the shock itself would likely present more sharply in non-thermal emission maps. In this scenario, energies of order 10 PeV may be attained in some regions of the bubble (based on equation~\ref{eq:emax_est}, but using the values appropriate for the shock), but energies of order 1 PeV would be more typical across most of the shock layer.}

While we do not invoke any particular acceleration mechanism to supply CRs, a choice of $E_{\rm max}$ between 10 TeV and 1,000 TeV covers the range of values suggested for advected CRs or those accelerated in the jets { or the bubble shock front.} 
As such, there is value in considering how our results differ when adopting an alternative upper limit to the hadronic CR spectrum, instead of our fiducial choice of $E_{\rm max} = 10\;\!{\rm TeV}$. Figure~\ref{fig:max_cutoff} shows the multi-wavelength bubble spectrum that results (at 1 Myr and 7 Myr) from setting $E_{\rm max} = 100$ TeV and 1,000 TeV, compared to the fiducial case. These show the case of a {hadronic} spectrum. 

Figure~\ref{fig:max_cutoff} reveals that the emission across much of the spectrum is increased with a higher choice of hadronic $E_{\rm max}$. This is due to the higher multiplicities associated with more energetic hadronic interactions, which supply secondary electrons over a broad range of energies. At radio frequencies, the synchrotron emission increases roughly in proportion to $E_{\rm max}$ at 1 Myr, and the position of the synchrotron peak also shifts {strongly to} higher frequencies. Conversely, at 7 Myr, the peak shifts quite noticeably to higher energies with incremental increases in $E_{\rm max}$, but the enhanced synchrotron emission overall is less {severe}. The reason for this is non-trivial, and lies in how the CR energy density and gas density (which mediates the hadronic interaction rate and, hence, the injection location of the secondary electrons) in the simulation are distributed compared to the magnetic field. At 1 Myr, when the draping layer is broadly intact near the outer shocked layers, much of the additional secondary CR injection is happening in the vicinity of stronger magnetic fields. However, the decayed structure of the bubble and the disconnect between the CR energy density and the surface layer by 7 Myr would diminish the strong relation between radio synchrotron intensity and $E_{\rm max}$. {At higher energies, in $\gamma$-rays, the inverse Compton emission} also transitions from CMB-dominated to ISRF-dominated at higher energies with increases in $E_{\rm max}$. This impacts very strongly on the $\gamma$-ray emission above a GeV, both at 1 Myr and 7 Myr, which becomes substantially brighter for $E_{\rm max} = 100$ or 1,000 TeV compared to the fiducial case.  

\subsubsection{Spectral shape}

Alternative CR spectral forms have been considered instead of the simple power law adopted in this work. For example,~\citet{Ackermann2014ApJ} find that a power law with an exponential cutoff spectrum gives a better fit to \textit{Fermi}-LAT data of the Galactic bubbles at high energies. 
Our exact choice of CR spectrum used initially for a leptonic bubble (if reasonable) is unlikely to be important. Radiative cooling quickly causes the spectrum to form a power-law with an exponential cut-off (see~\citetalias{OwenYang2021MNRAS}). However, the multi-wavelength emission from a hadronic/hybrid bubble would be more sensitive to the choice of spectral form assumed. As alternative models would usually vary the high energy part of the spectrum (e.g. by adopting an exponential cut-off, or broken power-law), the inverse Compton $\gamma$-ray emission would be most affected. Fewer CRs at higher energies would yield lower $\gamma$-ray emission above a TeV, or show reductions in emission intensity at lower energies from reduced secondary electron production. However, more detailed investigation of the impacts of the underlying CR spectrum on the emission signatures from a galaxy bubble is worthy of a dedicated work, and is left to a future study. 

\subsection{Observational considerations}

{Previously,~\citetalias{OwenYang2021MNRAS} investigated the detection prospects for leptonic bubbles around nearby Milky Way-like galaxies, of a similar configuration to the Galactic \textit{Fermi} bubbles. Host galaxies to distances of $\sim$ 20 Mpc were considered, where it was shown such bubbles would be detectable for at least} {1.9 Myr} {of their evolution with the up-coming SKA. Moreover, bubbles located within a distance of $\sim$ 15 Mpc were found to be detectable at radio frequencies for at least 7 Myr of their lifetimes. 
Based on instrument sensitivity alone, detection of nearby leptonic bubbles would be possible at $\gamma$-ray energies with current generation and up-coming facilities (e.g. \textit{Fermi}-LAT at energies between $\sim$ 20 MeV and $\sim$ 300 GeV and CTA between 0.1 and 1 TeV), but this would typically only be in the earlier stages of their evolution (up to ages of $\leq 1.5$ Myr).} 

{In Appendix~\ref{sec:obs_app} we extended this earlier analysis to also assess the detection prospects of CR bubbles with hybrid and hadronic compositions. Broadly, we found that leptonic bubbles are generally detectable for a greater period of their lifetime than their hadronic counterparts, and that radio observations offer the best prospect to detect both hadronic and leptonic galaxy bubbles (with other wavelengths being less encouraging). Bubble radio emission would remain accessible to SKA for much of their considered evolution, and would likely be spatially resolvable for at least a few Myr. Radio synchrotron emission has already been detected from bubbles around galaxies in previous work. While those structures are not be guaranteed to be the same phenomena as considered here, they do share certain similar characteristics, including the alignment of the extended radio emission with the minor axis of the host galaxy~\citep{Baum1993ApJ, Elmouttie1998MNRAS, Kharb2006ApJ} and, if present, their co-orientation with AGN jets in many spiral Seyfert galaxies where extended galactic-scale radio structures are common~\citep{Gallimore2006AJ}. If these structures are part of the galaxy bubble `family', they could represent systems at a different evolutionary stage and/or subject to different intensities of energy injection and persistence~\citep{Guo2012ApJ} than those considered in this work.}

\subsection{Additional remarks}
\label{sec:remarks}

Our results are sensitive to the physical properties of our simulated galaxy bubbles, and the numerical set up of our simulations and post-processing calculations. The physical properties are set by the initial conditions of our bubble simulations, which include the energy injected during the initial outburst and its duration~\citep[e.g.][]{Guo2012ApJ, Zhang2020ApJ}, the effect of multiple outbursts, and the ambient conditions surrounding the bubble. This study is the first that explores the multiwavelength non-thermal emission signatures of hadronic and leptonic CRs in galaxy bubbles, as well as their long term evolution. We thus put focus on the {similarities and differences between} hadronic and leptonic emission signatures in the current study, and leave detailed {investigation} of other physical bubble parameters {(e.g. AGN activity and variations in total energetics, host galaxy properties and bubble environment)} to future, dedicated works. {Even though we invoke a 
baseline model motivated by the Galactic \textit{Fermi} bubbles as a stage for our comparison, we consider that our results are still more generally useful. In particular, this work establishes information about the key spectral signatures and, hence, the most appropriate energy bands to probe the CR composition within bubbles around distant galaxies {(see Appendix~\ref{sec:obs_app} for further discussion)}. This would not substantially differ qualitatively over a range of model bubble parameter choices. }

In our numerical set-up, the spatial and spectral resolution adopted in our simulations and post-processing calculations is also consequential for our results. This work explored the evolution of galaxy bubbles over long periods (up to 7 Myr). Their expansion to large sizes ($\sim$ 40 kpc) also required large simulation volumes to be used. We found the computational requirements would be prohibitive to use the high resolutions adopted in bubble simulations in previous works (e.g.~\citealt{Yang2012ApJ, Yang2017ApJ}), which did not require as large volumes or as long evolutionary times. Thus, it was necessary to adopt a lower spatial resolution in our MHD simulations to moderate our computational requirements to an acceptable level. We also found that it was necessary to adopt a relatively low spectral resolution in our post-processing calculations (100 energy increments in constructing spectra, or integration over 10 energy increments in constructing the spatial emission maps). We found some emission processes to be particularly sensitive to the exact resolution of the simulation, especially those involving both the CR energy density and gas density {(hadronic interactions, and non-thermal bremsstrahlung)}. We therefore expect that future simulation developments at higher spatial and spectral resolutions will yield refined results -- particularly in regions near the top of the bubble. This would be of particular importance to the hadronic model, where low resolutions may act to artificially boost the production rate of secondary electrons in regions of high gas and CR energy density. The exact impact of these effects will be explored more thoroughly in future work.

\section{Conclusions}
\label{sec:section5}

In this work, we investigated the multiwavelength emission signatures from hadronic and leptonic processes in galaxy bubbles. We used 3D MHD simulations to model their development, after an initial 0.3 Myr outburst of energy and CRs from bipolar jets originating from the center of their host galaxy. We then post-processed the simulated bubbles, invoking hadronic, leptonic and hybrid composition scenarios to model their CR content. We determined their multi-wavelength non-thermal emission spectra, from radio frequencies through to {TeV} $\gamma$-rays, and also computed spatial emission maps in 5 energy bands (radio, X-ray, and $\gamma$-rays at {$\sim$sub-GeV, $\sim$GeV, and $\sim$TeV energies}). We compared the non-thermal emission properties of the bubbles under the different CR composition scenarios. This showed that, spatially, the emission at most energies more closely traced the underlying gas density of a hadronic or hybrid bubble than a leptonic one. {We also determined the differences in the emission spectra from hadronic, leptonic and hybrid bubbles, and considered how these could be detected in bubbles around external galaxies.} 

While constraints on a hadronic CR component in the Galactic \textit{Fermi} bubbles have established them to be likely lepton-dominated~\citep[e.g.][]{Abeysekara2017ApJ}, 
little is known about the composition of CRs in galaxy bubbles in general. There is no guarantee that bubbles around other galaxies would also harbor predominantly leptonic CRs, and investigating diverse bubble compositions will provide us with new insights about the characteristics of these phenomena. We have demonstrated that self-consistent multi-wavelength emission modeling of hadronic, leptonic and hybrid bubbles can provide information about differences in their emission signatures, and how such signatures would vary through the evolutionary progression of a bubble.  
Such efforts establish the possibility of inferring the compositions of bubbles observed around external galaxies with the suite of up-coming next-generation facilities that will operate across the electromagnetic spectrum in the coming decade. Together, these will offer a means to study the wider population and demographics of galaxy bubbles, and how our own Galaxy's \textit{Fermi} bubbles fit into this.

\section*{Data Availability}

The data generated in this research will be shared on reasonable request to the authors.

\section*{Acknowledgments}

ERO is supported by the Center for Informatics and Computation in Astronomy
  (CICA) at National Tsing Hua University (NTHU)
  through a grant from the Ministry of Education of Taiwan. HYKY acknowledges support from the Yushan Scholar Program of the Ministry of Education and the 
  National Science and Technology Council of Taiwan (NSTC 109-2112-M-007-037-MY3). We thank the National Center for High-performance Computing, Taiwan, for providing computational and storage resources, and the National Center for Theoretical Sciences, Taiwan, for provision of HPC time allocation, supported by a grant from NSTC (110-2124-M-002-012).
This work also used high-performance computing facilities at CICA, operated by the NTHU Institute of Astronomy. This equipment was funded by the Taiwan Ministry of Education and the Taiwan Ministry of Science and Technology. 
FLASH was developed in part by the DOE NNSA ASC- and DOE Office of Science ASCR-supported Flash Center for Computational Science at the University of Chicago. 
This research has made use of NASA's Astrophysics Data Systems. The authors thank the referees involved in the review of this article. Their critical, detailed comments led to substantial improvements in the manuscript.

\bibliographystyle{mnras} 
\bibliography{references} 


\appendix

\section{Simulation energetics}
\label{sec:appendixb}

\begin{figure}
\includegraphics[width=\columnwidth]{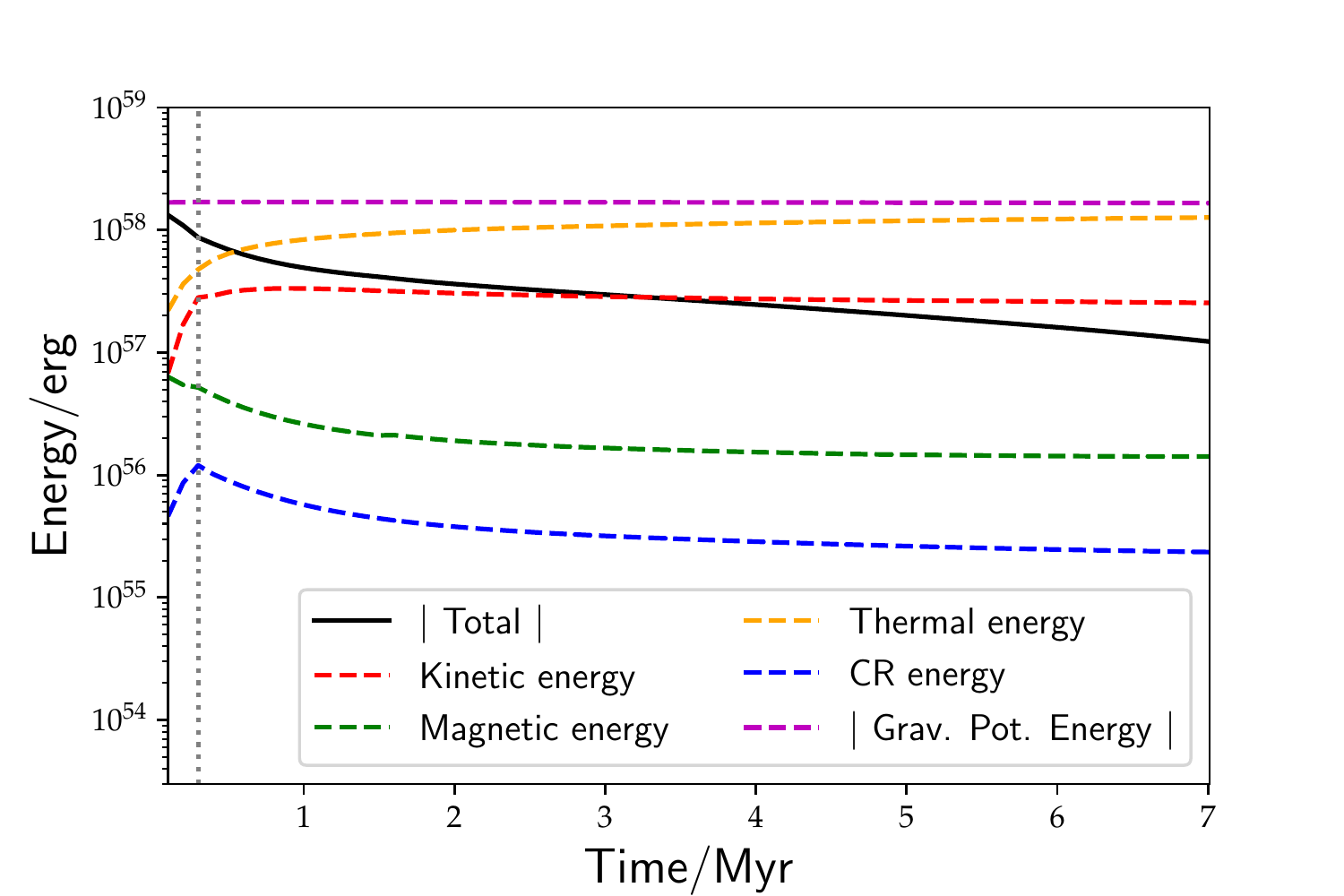}
\caption{{Total energy and its constituent components integrated over the full simulation box. The gravitational potential energy is a negative quantity, with its absolute value shown here for convenience. This dominates the energetics of the system at all times, thus the total energy is also shown as an absolute value and becomes less negative (increases) over time. The injected energy provision from the initial burst operates for the first 0.3 Myr of the simulation, with a cut-off time indicated by the vertical dashed grey line.}}
\label{fig:total_energy_simulation}
\end{figure}

{The evolution of the total energy in the simulation and its constituent components is plotted in Figure~\ref{fig:total_energy_simulation}. This shows an initial increase in CR energy, thermal energy and kinetic energy supplied by the 0.3 Myr energetic outburst, with a slow reduction in CR energy and magnetic energy at later times. The gravitational potential energy dominates the energetics of the simulation, but becomes comparable to the thermal energy at late times. The total energy (summed over all components) is a negative quantity, and its absolute value is shown by the black line in Figure~\ref{fig:total_energy_simulation}. While this increases (becomes less negative) at early times, a moderate late-time increase can also be seen. With the adopted diode boundary conditions used in our simulation, the total energy within the simulation 
domain is not expected to be conserved. As the halo gas is expanding due to the central injection, some gas would flow outside the simulation domain. This leads to an overall loss of energy. However, the expansion of the bubble also increases the gravitational potential energy of the whole system. This increase is greater than losses arising from gas leaving the simulation domain, so overall the total energy in the simulation increases with time.} 

\section{Observational considerations}
\label{sec:obs_app}

\subsection{Detection prospects}
\label{sec:external_bubbles}

\begin{table*}
\centering
{\begin{tabular}{*{12}{|lc|cc|cc|cc|cc|cc}}
 \multicolumn{2}{c}{} & \multicolumn{2}{c}{A: 8.3 - 15.3 GHz$^{c}$} & \multicolumn{2}{c}{B: 0.2 - 12 keV$^{c}$} & \multicolumn{2}{c}{C: 0.1 - 1 GeV} & \multicolumn{2}{c}{D: 1 - 10 GeV} & \multicolumn{2}{c}{E: 0.1 - 1 TeV} \\
\hline
{\bf Galaxy} & {\bf Distance/Mpc} & {\bf $t_{\rm A}$ / Myr}  & {$\theta_{\rm A}^{\rm max}$/~$^{\circ}$} & {\bf $t_{\rm B}$ / Myr}  & {$\theta_{\rm B}^{\rm max}$/~$^{\circ}$} & {\bf $t_{\rm C}$ / Myr} & {$\theta_{\rm C}^{\rm max}$/~$^{\circ}$} & {\bf $t_{\rm D}$ / Myr}  & {$\theta_{\rm D}^{\rm max}$/~$^{\circ}$} & {\bf $t_{\rm E}$ / Myr} & {$\theta_{\rm E}^{\rm max}$/~$^{\circ}$} \\
\hline
M31 & 0.770 & $>$7.0$^{a}$ & $>$5.5 & - & - & - & - & 1.5 & 1.9 & 1.4 & 1.6 \\
NGC 4710 & 2.14 & $>$7.0$^{a}$ & $>$2.0 & - & - & -  & - & \textit{0.80} & \textit{0.32} & 1.0 & 0.43 \\
NGC 3115 & 9.77 & $>$7.0$^{a}$ & $>$0.43 & - & - & - & - & - & -  & \textit{0.50} & \textit{0.047} \\
NGC 891 & 10.0 & $>$7.0$^{a}$ & $>$0.42 & - & - & - & - & - & - & \textit{0.50} & \textit{0.046}\\
NGC 4565 & 11.1 & $>$7.0$^{a}$ & $>$0.38 & - & - & - & - & - & - & \textit{0.40} & \textit{0.036} \\
NGC 7457 & 13.2 & $>$7.0$^{a}$ & $>$0.32 & - & - & -  & - & - & - & \textit{0.40} & \textit{0.030} \\
NGC 3877 & 14.1 & $>$7.0$^{a}$ & $>$0.30 & - & - & -  & - & - & - & \textit{0.30}$^{b}$ & \textit{0.020} \\
NGC 3198 & 14.5 & $>$7.0$^{a}$ & $>$0.29 & - & - & -  & - & - & - & \textit{0.30}$^{b}$ & \textit{0.020} \\
NGC 1386 & 15.3 & $>$7.0$^{a}$ & $>$0.28 & - & - & -  & - & - & - & - & -\\
NGC 5866 & 15.3 & $>$7.0$^{a}$ & $>$0.28 & - & - & -  & - & - & - & - & -\\
NGC 3079 & 16.5 & 4.6 & 0.18 & - & - & -  & - & - & - & - & -\\
NGC 4388 & 17.1 & 3.2 & 0.13 & - & - & -  & - & - & - & - & -\\
NGC 4526 & 17.2 & 3.2 & 0.13 & - & - & -  & - & - & - & - & -\\
NGC 7814 & 18.1 & 2.5 & 0.10 & - & - & -  & - & - & - & - & -\\
NGC 4013 & 18.9 & 2.2 & 0.091 & - & - & -  & - & - & - & - & -\\
NGC 4217 & 19.5 & 1.9 & 0.083 & - & - & -  & - & - & - & - & -\\
\hline 
\end{tabular}}
\caption{{\textbf{Leptonic bubble detectable lifetimes and angular sizes:} lifetimes over which galaxy bubbles would be observationally accessible in different bands (as indicated),   
if their configuration is the same as that adopted in our model and if they are located around nearby edge-on spiral and E/S0 type galaxies~\citep[see][]{Li2013MNRAS}. The selected galaxies are the same as those considered in~\citetalias{OwenYang2021MNRAS}. Sensitivities and angular resolutions of SKA (Band A, with angular resolution 0.09'', following from the lowest resolution indicated for the 8.3-15.3 GHz range; see~\citealt{Braun2019arXiv}), \textit{Athena} (Band B, angular resolution $\sim$ 5''), \textit{Fermi}-LAT (Bands C and D, with resolutions of 5$^{\circ}$ and 0.9$^{\circ}$ respectively based on the energy-dependent 68 per cent containment fraction) and CTA-North (Band E, resolution 0.13$^{\circ}$) are considered, with a 50-hour observation time (where appropriate) to determine observable lifetimes. Results for resolvable galaxies are shown in roman font, while those that would be theoretically detectable but not spatially resolvable with the considered instruments are shown in italics. Bubble sizes are estimated based on their extension at their maximum detectable age, with $\theta_{\rm max}$ computed from the distance and the full extent of symmetric North and South bubbles above and below the plane of their host galaxy.\\
\textbf{Notes}: \\
 $^{a}$ Ages greater than 7 Myr would presumably be observable for longer, but exceeded the evolution time of our simulations. These values are thus indicated as a lower limit. \\
 $^{b}$ For detectable lifetimes below 0.3 Myr, the bubble would still be brightening and subject to ongoing energy injection. We consider that these cases would not be detectable as bubbles. \\
 $^{c}$ Bands A and B are the same as considered in~\citetalias{OwenYang2021MNRAS}, and observable lifetimes therefore correspond to the values found in the earlier work (Bands C, D and E are defined differently).}}
\label{tab:observable_times_leptonic}
\end{table*} 

\begin{table*}
\centering
{\begin{tabular}{*{12}{|lc|cc|cc|cc|cc|cc}}
 \multicolumn{2}{c}{} & \multicolumn{2}{c}{A: 8.3 - 15.3 GHz} & \multicolumn{2}{c}{B: 0.2 - 12 keV} & \multicolumn{2}{c}{C: 0.1 - 1 GeV} & \multicolumn{2}{c}{D: 1 - 10 GeV} & \multicolumn{2}{c}{E: 0.1 - 1 TeV} \\
\hline
{\bf Galaxy} & {\bf Distance/Mpc} & {\bf $t_{\rm A}$ / Myr}  & {$\theta_{\rm A}^{\rm max}$/~$^{\circ}$} & {\bf $t_{\rm B}$ / Myr}  & {$\theta_{\rm B}^{\rm max}$/~$^{\circ}$} & {\bf $t_{\rm C}$ / Myr} & {$\theta_{\rm C}^{\rm max}$/~$^{\circ}$} & {\bf $t_{\rm D}$ / Myr}  & {$\theta_{\rm D}^{\rm max}$/~$^{\circ}$} & {\bf $t_{\rm E}$ / Myr} & {$\theta_{\rm E}^{\rm max}$/~$^{\circ}$} \\
\hline
M31 & 0.770 & $>$7.0 & $>$5.5 & 1.1 & 1.4 & \textit{0.5} & \textit{0.6} & 1.5 & 1.9 & 1.9 & 2.1 \\
NGC 4710 & 2.14 & $>$7.0 & $>$2.0 & - & - & -  & - & \textit{0.80} & \textit{0.32} & 1.0 & 0.43 \\
NGC 3115 & 9.77 & 6.5 & 0.41 & - & - & - & - & - & -  & - & - \\
NGC 891 & 10.0 & 6.5 & 0.40 & - & - & - & - & - & - & - & - \\
NGC 4565 & 11.1 & 5.7 & 0.33 & - & - & - & - & - & - & - & - \\
NGC 7457 & 13.2 & 4.3 & 0.22 & - & - & -  & - & - & - & - & - \\
NGC 3877 & 14.1 & 3.4 & 0.17 & - & - & -  & - & - & - & - & - \\
NGC 3198 & 14.5 & 3.2 & 0.16 & - & - & -  & - & - & - & - & - \\
NGC 1386 & 15.3 & 2.8 & 0.13 & - & - & -  & - & - & - & - & -\\
NGC 5866 & 15.3 & 2.5 & 0.12 & - & - & -  & - & - & - & - & -\\
NGC 3079 & 16.5 & 2.3 & 0.11 & - & - & -  & - & - & - & - & -\\
NGC 4388 & 17.1 & 2.2 & 0.10 & - & - & -  & - & - & - & - & -\\
NGC 4526 & 17.2 & 2.1 & 0.10 & - & - & -  & - & - & - & - & -\\
NGC 7814 & 18.1 & 2.1 & 0.092 & - & - & -  & - & - & - & - & -\\
NGC 4013 & 18.9 & 2.1 & 0.088 & - & - & -  & - & - & - & - & -\\
NGC 4217 & 19.5 & 2.0 & 0.084 & - & - & -  & - & - & - & - & -\\
\hline 
\end{tabular}}
\caption{{\textbf{Hadronic bubble detectable lifetimes and angular sizes:} same as Table~\ref{tab:observable_times_leptonic}, but for hadronic CR bubble compositions.}}
\label{tab:observable_times_hadronic}
\end{table*} 

In Tables~\ref{tab:observable_times_leptonic} and~\ref{tab:observable_times_hadronic}, we assess the detection prospects of CR bubbles, if they were located around nearby galaxies. In particular, we estimate the age to which hadronic and leptonic bubbles would remain visible in the five energy bands considered previously in section~\ref{sec:broadband_emission_maps} (radio in Band A: 8.3 - 15.3 GHz, X-rays in Band B: 0.2 - 12 keV, and $\gamma$-rays in Bands C: 0.1-1 GeV, D: 1-10 GeV and E: 0.1 - 1 TeV). We adopt sensitivities of four current/up-coming instruments appropriate for each of the five bands (namely, SKA\footnote{{See~\cite{Braun2019arXiv}.}} for Band A, \textit{Athena}\footnote{{See~\cite{Nandra2013arXiv}.}} for Band B, \textit{Fermi}-LAT\footnote{\url{https://fermi.gsfc.nasa.gov}} for Bands C and D, and CTA\footnote{{Available online:} \url{https://www.cta-observatory.org/science/ctao-performance/}} for Band E) and calculate the time period these instruments could detect our simulated bubble, if it was located at the positions of selected nearby edge-on Milky Way-sized spiral and E/S0 type galaxies within 20 Mpc~\citep[see][]{Li2013MNRAS}. Galaxies fitting this criteria would be reflective of systems where possible bubble structures aligned along the minor axis of their host have been identified previously in various wavebands~\citep[e.g.][]{Gallimore2006AJ}, and so would seem to be most suitable as potential hosts for the purposes of our analysis.\footnote{This sample is the same as that considered in~\citetalias{OwenYang2021MNRAS}, where the computed detectable timescales in Bands A and B are the same. Bands C, D and E are specified differently here, hence the differences in the observable lifetimes quoted in Table~\ref{tab:observable_times_leptonic} compared to the previous work.}

A further aspect concerning the detection prospects for galaxy bubbles is their spatial size, and whether this would be resolvable by instruments in each band. Practically, only those bubbles for which instruments have both sufficient sensitivity and angular resolution to discern could plausibly be considered to be within observational reach. In Tables~\ref{tab:observable_times_leptonic} and~\ref{tab:observable_times_hadronic} we therefore also indicate the maximum angular size of the bubbles expected around each host galaxy, computed according to our model and corresponding to their extent at the latest age they could be detected. We compare this to the nominative angular resolution of the four instruments in each band. Where detections are not possible at all, no value for maximum bubble detection age is shown. For systems that would be sufficiently bright but not sufficiently large to be resolved, the maximum detectable age and angular extent are shown in italics. While these are included for completeness and to illustrate the evolution of the luminosity of hadronic and leptonic bubbles around external galaxies, we consider these spatially-unresolved cases would not truly be detectable.

We find that radio observations offer the best prospect for the detection of both hadronic and leptonic galaxy bubbles. Tables~\ref{tab:observable_times_leptonic} and~\ref{tab:observable_times_hadronic}
show that their synchrotron emission would be accessible for much of a bubble's lifetime with SKA, regardless of CR composition. Moreover, bubbles would likely be spatially resolvable for at least a few Myr in the radio band, even out to distances of nearly 20 Mpc (cf. NGC 4217). 
It can also be seen that leptonic bubbles are typically visible for a greater period of their lifetime than their hadronic counterparts. This may even be understated in Table~\ref{tab:observable_times_leptonic}, where our simulations only allowed us to estimate observable lifetimes up to 7 Myr. It is plausible that some of the more local systems would be detectable for much longer than this (or equivalently would be observed to be much brighter), and would seem to "outshine" their hadronic counterparts at these radio frequencies.

Detection prospects of both hadronic and leptonic bubbles in other bands are comparatively poor, with only bubbles around M31 likely within reach of \textit{Fermi}-LAT in $\gamma$-rays,\footnote{{We note that this prediction is consistent with the possible detection of bubbles around M31~\citep{Pshirkov2016MNRAS}, which may have a slightly younger age than the Galactic \textit{Fermi} bubbles.}} and only a hadronic bubble offering plausible detection prospects in X-rays (even when considering the improved expected sensitivity of \textit{Athena}), where the inverse Compton emission component peaks at lower energies than the leptonic model. However, we note that the inclusion of thermal bremsstrahlung emission from bubbles in future work is likely to reveal substantially more promising X-ray detection prospects than our results here would suggest.

For higher-energy $\gamma$-rays, pion decays in hadronic and hybrid bubbles would lead to more stable 0.1-1 TeV emission than their leptonic counterparts for longer periods of their evolution. 
For example, a leptonic bubble located around M31 would remain visible (to CTA) for the first 1.4 Myr of its lifetime in the 0.1-1 TeV energy band. By contrast, a hadronic bubble at the same distance would remain detectable up to 1.9 Myr in this band. 
The $\gamma$-ray luminosity of both leptonic and hadronic bubbles in their early stages would, however be similar. In terms of accessible distances, both leptonic and hadronic bubbles would be detectable around possible nearby hosts galaxies out to $\sim 2$ Mpc (unless the maximum hadronic CR energy were substantially higher than assumed in this work; cf. Figure~\ref{fig:max_cutoff}).

\subsection{Distinguishing hadronic and leptonic bubbles}
\label{sec:had_lep_bubbles}

After their discovery, it was quickly established that hadronic and leptonic scenarios for the Galactic \textit{Fermi} bubbles would have very different natural timescales (see, e.g.~\citealt{Ackermann2014ApJ}). Thus, to distinguish between compositions, it has become necessary to independently determine their age. This would also be the case for bubbles around external galaxies, where data would be poorer.  
While the non-thermal luminosity of bubbles is detectable and also differs according to their CR composition 
(comparing Band A in Tables ~\ref{tab:observable_times_leptonic} and~\ref{tab:observable_times_hadronic}), this information alone is not sufficient to resolve their age or to determine whether they are leptonic or hadronic, and further constraints are required. 
Moreover, our simulations only represent one possible bubble configuration, and a thorough exploration of model parameters is necessary to put forward any practical means of distinguishing between hadronic and leptonic bubbles reliably. However, we consider that our results may still be used to consider a possible operational procedure, and as a basis to indicate how bubbles may be distinguished when a more robust theoretical understanding of these systems and their variation is available.  

Figure~\ref{fig:total_energy_simulation} shows that the bubbles in our simulations are thermally-dominated.
Thus, if an understanding of how the partition between different energy components in a bubble evolves can be reasonably modeled (cf. Figure~\ref{fig:total_energy_simulation}), knowledge of both the average gas temperature and non-thermal luminosity of a bubble together would allow a constraint to be placed the ratio of thermal and CR energy densities, i.e. $\mathcal{E}(t) = e_{\rm th}(t)/e_{\rm CR}(t)$.\footnote{Previous work used the \ion{O}{vii}/\ion{O}{viii} line ratio to obtain the gas temperature of the \textit{Fermi} bubbles in order to constrain scenarios of their origins~\citep{Sarkar2017MNRAS}. Tuning against the Galactic \textit{Fermi} bubbles could 
resolve the degeneracy between the non-thermal luminosity and energy content in the CRs to allow $e_{\rm CR}(t)$ to be obtained (cf. section~\ref{sec:post_proc} and the $f_{\rm emit}$ parameter).} This quantity increases monotonically as a bubble ages, after the initial burst of energy injection. Thus, it can allow for an unambiguous determination of a bubble's age (or, at least, the time elapsed since it was last subject to a substantial injection of energy). Once the age is known, the Band A luminosity would then allow the CR composition of a bubble to be estimated. Variation in the evolution of the ratio $\mathcal{E}(t)$ with alternative total bubble energetics is left for exploration in dedicated follow-up work.

\bsp	
\label{lastpage}
\end{document}